\documentclass[10pt,final,journal]{IEEEtran}
\usepackage[switch]{lineno} 
\usepackage{tabularx} 
\usepackage{booktabs}
\usepackage{multirow}
\usepackage{makecell}

\usepackage{stmaryrd}
\usepackage{graphicx}
\usepackage{amsthm} 
\usepackage{amsmath} 
\usepackage{bm}
\usepackage{graphicx}
\usepackage{epstopdf}
\usepackage{color}
\usepackage{float}
\usepackage[colorlinks=false,linkcolor=black,urlcolor=black,bookmarksopen=true, hidelinks]{hyperref}
\usepackage{bookmark}
\usepackage[flushleft]{threeparttable}
\usepackage{stackengine}
\usepackage{scalerel}
\usepackage{mathtools}

\usepackage{xcolor,soul}

\usepackage{cite}

\ifCLASSINFOpdf
\else
\fi
\usepackage{amsfonts,amssymb}

\usepackage{amsmath}
\usepackage{algorithm}
\usepackage{algpseudocode}
\ifCLASSOPTIONcompsoc
\usepackage[caption=false,font=normalsize,labelfont=sf,textfont=sf]{subfig}
\else
\usepackage[caption=false,font=footnotesize]{subfig}
\fi
\begin{document}
%
\title{Sequential Doppler Shift based Optimal Localization and Synchronization with TOA}

\author{Sihao~Zhao, \textit{Member, IEEE},
	Ningyan~Guo,
	Xiao-Ping~Zhang, \textit{Fellow, IEEE},
	Xiaowei~Cui,
	and~Mingquan~Lu
	\thanks{This work was supported in part by the Natural Sciences and Engineering Research Council of Canada (NSERC), Grant No. RGPIN-2020-04661. \textit{(Corresponding author: Xiao-Ping Zhang.)}}
	\thanks{S. Zhao, X.-P. Zhang are with the Department of Electrical, Computer and Biomedical Engineering, Ryerson University, Toronto, ON M5B 2K3, Canada (e-mail: sihao.zhao@ryerson.ca; xzhang@ee.ryerson.ca).}
	\thanks{N. Guo, X. Cui are with the Department of Electronic Engineering,
		Tsinghua University, Beijing 100084, China (e-mail: guoningyan@tsinghua.edu.cn; cxw2005@tsinghua.edu.cn).}
	\thanks{M. Lu is with the Department of Electronic Engineering,
		Beijing National Research Center for Information Science and Technology, Tsinghua University, Beijing 100084, China. (e-mail: lumq@tsinghua.edu.cn).}
}

\markboth{}%
{Shell \MakeLowercase{\textit{et al.}}: Bare Demo of IEEEtran.cls for IEEE Journals}
%




\maketitle

\begin{abstract}
Doppler shift is an important measurement for localization and synchronization (LAS), and is available in various practical systems.
Existing studies on LAS techniques in a time division broadcast LAS system (TDBS) only use sequential time-of-arrival (TOA) measurements from the broadcast signals.
In this paper, we develop a new optimal LAS method in the TDBS, namely LAS-SDT, by taking advantage of the sequential Doppler shift and TOA measurements. It achieves higher accuracy compared with the conventional TOA-only method for user devices (UDs) with motion and clock drift. Another two variant methods, LAS-SDT-v for the case with UD velocity aiding, and LAS-SDT-k for the case with UD clock drift aiding, are developed. We derive the Cram\'er-Rao lower bound (CRLB) for these different cases. We show analytically that the accuracies of the estimated UD position, clock offset, velocity and clock drift are all significantly higher than those of the conventional LAS method using TOAs only.
Numerical results corroborate the theoretical analysis and show the optimal estimation performance of the LAS-SDT.
	
\end{abstract}



%
\IEEEpeerreviewmaketitle

\section{Introduction}\label{Introduction}
%
%
%
%
\IEEEPARstart{L}{o}CALIZATION and synchronization (LAS) techniques for user devices (UDs) using a set of measurements has gained significant attention in a variety of applications, such as Internet of Things (IoT), emergency rescue, aerial surveillance, Internet of Vehicles (IoV) and target detection and tracking \cite{kuutti2018survey,yang2014overview,ferreira2017localization,beard2002coordinated}. To achieve LAS for moving UDs with clock drift, there are usually several synchronous anchor nodes (ANs) at known locations to transmit signals that the moving UDs capture to obtain measurements for LAS. Typical measurements include time-of-arrival (TOA), time difference of arrival (TDOA), received signal strength (RSS), angle of arrival (AOA) and their combinations \cite{guvenc2009survey,shao2014efficient,tomic20163,wang2012novel,hu2017robust,zhao2021semidefinite,hamer2018self,zhao2020closed}, since these measurements are associated with the moving UDs' position and timing information.

Doppler shift is an important measurement for LAS in addition to the measurements mentioned above. It is adopted by many real-world LAS systems such as the global navigation satellite systems (GNSSs). Unlike other measurements, Doppler shifts are directly related to the UD velocity and clock drift and thus can achieve LAS solely or improve LAS accuracy combined with other measurements.
Using the Doppler shifts only or along with the other measurements, including TOA, TDOA and AOA, etc., to estimate the source or the target localization and velocity is widely studied in literature.

The studies in \cite{amar2008localization,tirer2017high,jinzhou2012linear,lee2007doppler,nguyen2018closed,gong2020auv,ahmed2020localization,deng2018doppler} employ the Doppler shifts alone to achieve position and velocity estimation for a stationary or moving target or source. Various estimation methods are investigated in these works. The direct position determination (DPD) for locating a single emitter is presented in \cite{amar2008localization,tirer2017high}. The Gauss-Newton iterative method using concurrent Doppler shift measurements from several satellites \cite{jinzhou2012linear} or sequential Doppler shifts from a moving unmanned aerial vehicle (UAV) \cite{lee2007doppler} are proposed. Closed-form methods for locating one or more stationary sources by different number of moving or stationary sensors are presented in  \cite{nguyen2018closed,gong2020auv,ahmed2020localization}. Semi-definite relaxation (SDR) methods for localization are developed \cite{ahmed2020localization,deng2018doppler}.

Utilizing Doppler shifts along with TOA or AOA measurements for LAS are investigated in \cite{jia2020localization,yang2016moving,yin2017direct,ho2007source,ho2004accurate,wei2009multidimensional,musicki2009mobile,wang2012semidefinite,wang2016efficient}. Localization for a moving target using time delay and Doppler shift measurements in the presence of sensors' motion are studied in \cite{jia2020localization}.
In \cite{yang2016moving}, the Doppler shifts and the elliptic distance measurements are explored to jointly estimate a moving target's position and velocity and a closed-form method is proposed. Localization for multiple stationary transmitters that uses angle and Doppler shift measurements is performed using a direct position determination (DPD) method based on the weighted subspace fitting (WSF) algorithm \cite{yin2017direct}. The studies in \cite{ho2007source,ho2004accurate,wei2009multidimensional,musicki2009mobile,wang2012semidefinite,wang2016efficient} exploit TDOA and frequency differences of arrival (FDOA) measurements to achieve source or target localization. The work in \cite{ho2007source} focuses on solving the localization problem in the presence of receiver random error. A closed-form solution, which does not need the initial guess and has low computational complexity, to determine the position and velocity of a moving target is proposed in \cite{ho2004accurate}. 
A multidimensional scaling (MDS) based method,
which is shown to be robust to the large measurement noise is presented in \cite{wei2009multidimensional}. The study in \cite{musicki2009mobile} proposes an extension of Gaussian mixture presentation of measurements-integrated tracking splitting (GMM-ITS) algorithm to track a moving emitter. SDR based methods for moving source localization with different formulations are proposed in \cite{wang2012semidefinite,wang2016efficient}.

However, all of the above-mentioned studies assume that the ranging signals to form TOA measurements are sent or received at the same time, and the UD does not have clock drift. This assumption restricts their applications a time division broadcast system (TDBS), which utilizes sequential TOA measurements for LAS, and leads to LAS errors caused by the UD clock drift.

In a TDBS, ANs periodically broadcast the signals according to their pre-scheduled launch time slots and the UDs passively receive the signals to obtain sequential measurements. Since they have separate time slots to transmit signals, the ANs in a TDBS do not need special radio frequency front end design to isolate the transmission and reception signals. This also brings an easier synchronization between ANs since each AN can receive the signals from other ANs to determine its own clock offset \cite{shi2019blas}. On the contrary, the concurrent measurement systems, such as Global Positioning System (GPS) and the global navigation satellite system (GLONASS), require complex pseudorandom codes or need complex radio frequency front end design for different narrow frequency bands on each satellite to avoid interference, and have a high cost for inter-satellite synchronization \cite{misra2006global}. In addition, this TDBS scheme supports unlimited number of UDs and offers high safety for the UDs from being detected. Due to these benefits, the TDBS has attracted more attention and the LAS techniques in such a system have been widely studied recently \cite{hamer2018self,shi2019blas,dwivedi2013cooperative,zachariah2013self,carroll2014demand,yan2017asynchronous,zachariah2014schedule}. Many of the prior works \cite{dwivedi2013cooperative,zachariah2013self,carroll2014demand,yan2017asynchronous,zachariah2014schedule} focus on resolving the UD localization problem and do not address the issues of clock synchronization and velocity estimation. Methods for jointly estimating the UD position and clock parameters using sequential TOA measurements in a TDBS are proposed in \cite{hamer2018self,shi2019blas}. However, these studies ignore the UD's motion during the reception period for multiple measurements, making the methods not applicable for moving UDs.

There are some recent studies on the LAS problem in a TDBS for UDs with motion, clock offset and clock drift. Zhao et al. \cite{zhao2020optloc} develop a set of optimal localization methods for the moving UDs in different cases, and analyze their performances. Shi et al. \cite{shi2020sequential} propose a two-step weighted least squares (WLS) method to jointly estimate the position, velocity and clock parameters of the UDs in the presence of position uncertainties of the ANs. Guo et al. \cite{guo2021new} propose a closed-form LAS approach for moving UDs in a TDBS with synchronous ANs at known positions. However, they only use the sequential TOA measurements, and the utilization of Doppler shifts is not studied.

Doppler shifts are associated with the relative motion and clock drift between the ANs and UDs. Utilizing the Doppler shift measurements in a TDBS has the potential to improve the LAS performance for a moving UDs with clock drift. Yet, there is no report on LAS methods in a TDBS using Doppler shift and TOA measurements, not to mention the studies on performance evaluation.

In this paper, taking advantage of the Doppler shift to achieve high-accuracy LAS for UDs with motion and clock drift in the TDBS, we  develop a new optimal LAS method using the sequential Doppler and TOA measurements, namely LAS-SDT. We first formulate the LAS problem as a maximum likelihood (ML) estimator and develop an iterative algorithm to solve it. In special cases, where aiding information such as the UD velocity from external sensors and the UD clock drift from beforehand calibration are available, we propose two variant algorithms, i.e., LAS-SDT-v for the case with UD velocity aiding information, and LAS-SDT-k for the case with clock drift aiding information. We analyze the estimation errors of the LAS-SDT and derive the CRLB. We show that the estimation accuracy of the LAS-SDT is higher than that of the conventional TOA-only method.
We demonstrate analytically that the LAS error increases when the aiding information of velocity or clock drift deviates from the true value. Simulation results show the superior estimation accuracy of the new LAS-SDT over the conventional TOA-only method, and validate all the theoretical analyses.

The rest of the paper is organized as follows. In Section II, we present the system model and formulate the problem. We develop the optimal LAS method using both sequential Doppler shift and TOA measurements in Section III. The LAS estimation performances in different cases are analyzed in Section IV. Section V presents the performance evaluation based on numerical simulations. Finally, Section VI draws the conclusion of this paper.

Main notations are summarized in Table \ref{table_notation}.

\begin{table}[!t]
	\caption{Notation List}
	\label{table_notation}
	\centering
	\begin{tabular}{l p{5.5cm}}
		\toprule
		lowercase $x$&  scalar\\
		bold lowercase $\boldsymbol{x}$ & vector\\
		bold uppercase $\bm{X}$ & matrix\\
		$\hat{x}$, $\hat{\boldsymbol{x}}$ & estimate of a variable\\
		$\Vert \boldsymbol{x} \Vert$ & Euclidean norm of a vector\\
		$\Vert \boldsymbol{x}\Vert _{\bm{W}}^2$ & square of Mahalanobis norm, i.e., $\boldsymbol{x}^T\bm{W}\boldsymbol{x}$\\
		$\mathrm{tr}(\bm{X})$ & trace of a matrix\\
		$\left|\bm{X}\right|$ & determinant of a matrix\\
		$[\bm{X}]_{u,:}$, $[\bm{X}]_{:,v}$ &the $u$-th row and the $v$-th column of a matrix, respectively\\
		$[\bm{X}]_{u:v,m:n}$ &sub-matrix with the $u$-th to the $v$-th rows and the $m$-th to the $n$-th columns\\
		$[\bm{X}]_{u,v}$ &entry at the $u$-th row and the $v$-th column of a matrix\\
		$[\boldsymbol{x}]_{u}$ &the $u$-th element of a vector\\
		$\mathbb{E}[\cdot]$ & expectation operator \\
		$\mathrm{diag}(\cdot)$ & diagonal matrix with the elements inside\\
		$M$ & number of ANs\\
		$N$ & dimension of all the position and velocity vectors, i.e., $N=2$ in 2D case and $N=3$ in 3D case\\
		$i$, $j$ & index of the measurements \\
		$\boldsymbol{0}_{N}$, $\boldsymbol{1}_{N}$ & $N$-element vectors with all zeros and ones\\
		$\bm{O}_{M \times N}$, $\bm{O}_{N}$  & $M \times N$, and $N \times N$ matrices with all-zero entries\\
		$\bm{I}_{N}$ &$N \times N$ identity matrix\\
		$\boldsymbol{q}_{i}$ & known position vector of AN \#$i$\\
		$\boldsymbol{p}$, $\boldsymbol{v}$ &  position and velocity vector of UD\\
		$b$, $k$ &  clock offset and clock drift between UD and ANs\\
		$\rho_{i}$ & TOA measurement between UD and AN \#$i$\\
		$d_i$ & Doppler shift measurement between UD and AN \#$i$\\
		$\boldsymbol{e}$ & unit line-of-sight (LOS) vector from the UD to the AN\\
		$\Delta t$, $\Delta t_i$ &  time interval between successive measurements, and $\Delta t_i=\Delta t \cdot (i-1)$\\
		$\boldsymbol{\theta}$ &  parameter vector\\
		$\varepsilon$, $\sigma^2$ &  Gaussian measurement noise and variance\\
		$\mathcal{F}$ &  Fisher information matrix\\
		$\bm{W}$ & weighting matrix\\
		$\bm{\Sigma}_{d}$, $\bm{\Sigma}_{\rho}$, & variance matrices of Doppler shift and TOA noises\\
		$\bm{G}$ & design matrix\\
		$\mu$& estimation bias\\
		$\bm{Q}$ & estimation error variance matrix\\		
		\bottomrule
	\end{tabular}
\end{table}

\section{Problem Formulation} \label{problem}
\subsection{Time Division Broadcast LAS System (TDBS) Model}
\begin{figure}
	\centering
	\includegraphics[width=0.94\linewidth]{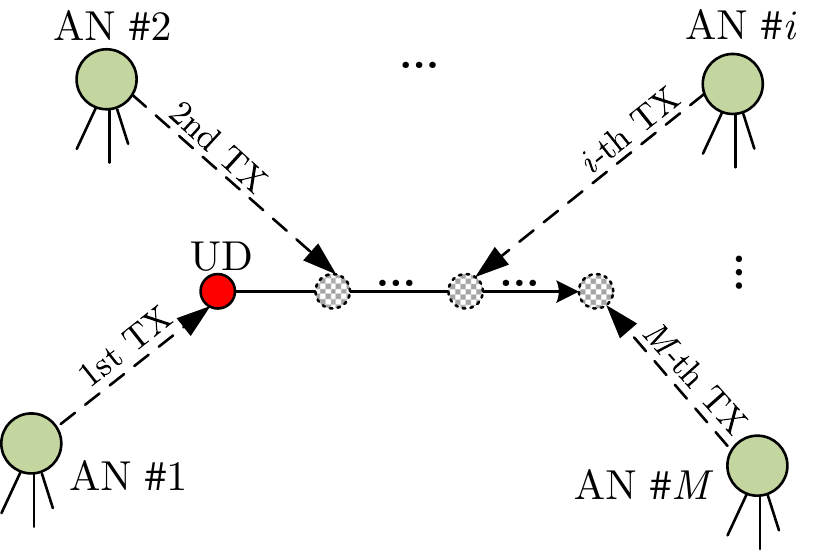}
	\caption{ANs and UD in a TDBS. The ANs transmit signals in a sequential manner. The UD obtains the Doppler shift and TOA measurements from the received signals. The position and clock offset of the moving UD change with time.
	}
	\label{fig:sysfig}
\end{figure}

We consider a TDBS, in which $M$ ANs broadcast signal, and moving UDs receive this signal and obtain the Doppler shift and TOA measurements, as shown in Fig. \ref{fig:sysfig}. We denote the known position of the $i$-th AN by $\boldsymbol{q}_i$, where $i=1,\cdots,M$. All ANs are synchronous and their time is denoted by $t$. They sequentially broadcast signals in non-overlap time slots periodically, e.g., AN \#1, AN \#2, ..., AN \#$M$ and then AN \#1. Without loss of generality, the sequential Doppler shift and TOA measurements in one broadcast round from all $M$ ANs are used.

A moving UD in the TDBS receives the signals from ANs to form Doppler shift and TOA measurements. The interval between successive measurements is assumed identical and denoted by $\Delta t$. We denote the unknown position and clock offset of a UD at the beginning of one broadcast round by $\boldsymbol{p}$ and $b$, respectively. The UD velocity and clock drift are denoted by $\boldsymbol{v}$ and $k$, respectively. The position and velocity of both the ANs and UD are of $N$ dimensional, and $N=2$ or $N=3$. Since one broadcast round of the ANs has a very short time, the velocity and clock drift of the UD are treated as constant during this short period. We aim to obtain the unknown parameters of a UD in the TDBS using the Doppler shift and TOA measurements from the sequential broadcast signals within one broadcast round. The Doppler and TOA  measurements are modeled in the next subsection.

\subsection{Doppler Shift and TOA Measurement Model}
The Doppler shift measurement with respect to AN \#$i$, denoted by $d_i$, is obtained by comparing the frequencies of the received signal and the local oscillator \cite{kaplan2005understanding}. It is related to the UD's velocity and clock drift. We divide the Doppler measurement by the carrier frequency and then multiply the signal propagation speed to convert it from frequency to velocity. It is then expressed by
\begin{equation} \label{eq:Doppleri}
	d_i =-\boldsymbol{v}^T \frac{\boldsymbol{q}_i - \boldsymbol{p}-\boldsymbol{v}\cdot \Delta t_i}{\Vert \boldsymbol{q}_i - \boldsymbol{p}-\boldsymbol{v}\cdot \Delta t_i\Vert} + k + \varepsilon_{d_i} \text{,}\; i=1,\cdots,M\text{,}
\end{equation}
where $\Delta t_i= \Delta t \cdot (i-1)$ is the time interval between the start time of one broadcast round and the reception time of the broadcast signal from AN \#$i$, $\varepsilon_{d_i}$ is the measurement noise, which follows an independent zero-mean Gaussian distribution with a variance of $\sigma_{d_i}^2$, i.e., $\varepsilon_{d_i} \sim \mathcal{N}(0,\sigma_{d_i}^2)$.

By differencing the local reception timestamp and the local transmission timestamp, we can obtain the TOA measurement at the UD. Since it is equivalent to a range when we multiply the signal propagation speed, it is also referred to as pseudorange in the literature \cite{misra2006global,kaplan2005understanding}. Following the measurement model in \cite{zhao2020optloc,zhao2021semidefinite}, the sequential TOA measurement from AN \#$i$, denoted by $\rho_i$, is written by
\begin{equation} \label{eq:rhoi}
	\rho_i =\Vert \boldsymbol{q}_i - \boldsymbol{p}-\boldsymbol{v}\cdot \Delta t_i \Vert + b + k \cdot \Delta t_i + \varepsilon_{\rho_i} \text{,}\; i=1,\cdots,M\text{,}
\end{equation}
where $\varepsilon_{\rho_i}$ is the measurement noise, following an independent zero-mean Gaussian distribution with a variance of $\sigma_{\rho_i}^2$, i.e., $\varepsilon_{\rho_i} \sim \mathcal{N}(0,\sigma_{\rho_i}^2)$, and the time-related terms $\rho_i$, $b$ and $k\Delta t_i$ are all converted to distances by multiplying the signal propagation speed and have the unit of meter.


The LAS problem for a moving UD with clock drift in such a TDBS is to estimate the UD position $\boldsymbol{p}$ and the clock offset $b$ using $M$ Doppler shifts and $M$ TOA measurements given by (\ref{eq:Doppleri}) and (\ref{eq:rhoi}). This problem is important in nowadays LAS applications and has not been studied in the literature. We will develop an optimal LAS method in the next section.


\section{Optimal LAS for UD with Motion and Clock Drift in TDBS}
As modeled in Section \ref{problem}, the unknown parameters of interest include position $\boldsymbol{p}$ and clock offset $b$. We investigate three practical cases, i) no prior knowledge on any parameters, ii) with velocity aiding information, and iii) with clock drift aiding information. The optimal LAS methods for the three cases are proposed in this section.

\subsection{Optimal Estimation for All Unknown Parameters}
\subsubsection{ML Estimator}

The parameter vector to be estimated is denoted by $\boldsymbol{\theta}$. Without any prior knowledge on any parameter, $\boldsymbol{\theta}$ contains the UD position, clock offset, velocity and clock drift, i.e.,
$$
\boldsymbol{\theta}=\left[\boldsymbol{p}^T,b,\boldsymbol{v}^T,k\right]^T \text{.}
$$

The relationship between the measurements and the unknown parameters is given by
\begin{equation} \label{eq:allmeasurements}
	\boldsymbol{\tau}=\left[\boldsymbol{d}^T, \boldsymbol{\rho}^T\right]^T = \mathit{h}(\boldsymbol{\theta}) + \boldsymbol{\varepsilon} \text{,}
\end{equation}
where $\boldsymbol{d}$ and $\boldsymbol{\rho}$ are the collective form of the Doppler shift measurement $d_i$ and the TOA measurement $\rho_i$, respectively, i.e., $\boldsymbol{d}=[d_1,\cdots,d_M]^T$ and $\boldsymbol{\rho}=[\rho_1,\cdots,\rho_M]^T$, $\boldsymbol{\varepsilon}$ is the collective form of all the measurement noises, i.e., $\boldsymbol{\varepsilon}=[\varepsilon_{d_1},\cdots,\varepsilon_{d_M},\varepsilon_{\rho_1},\cdots,\varepsilon_{\rho_M}]^T$, the Doppler and TOA measurement noise variances are denoted by $\bm{\Sigma}_{d}$ and $\bm{\Sigma}_{\rho}$, respectively, 
\begin{align}
	\bm{\Sigma}_{d}&=\mathrm{diag}\left(\sigma_{d_1}^2,\cdots, \sigma_{d_M}^2\right),\nonumber\\
	\bm{\Sigma}_{\rho}&=\mathrm{diag}\left(\sigma_{\rho_1}^2,\cdots, \sigma_{\rho_M}^2\right),
\end{align}
$\mathrm{diag}(\cdot)$ is a diagonal matrix comprised of the elements inside, and the function $h(\cdot)$ is a nonlinear function, which has a collective form as given by \eqref{eq:Doppleri} and \eqref{eq:rhoi}, i.e.,

\begin{align} \label{eq:funtheta}
	&\left[h(\boldsymbol{\theta})\right]_{i} = \nonumber\\
	&\left\{
	\begin{matrix}
		-\boldsymbol{v}^T \frac{\boldsymbol{q}_{i} - \boldsymbol{p}-\boldsymbol{v}\cdot \Delta t_{i}}{\Vert \boldsymbol{q}_{i} - \boldsymbol{p}-\boldsymbol{v}\cdot \Delta t_{i}\Vert} + k,\\
		\hspace{3.2cm}i=1,\cdots,M,\\
		\Vert \boldsymbol{q}_{i-M} - \boldsymbol{p}-\boldsymbol{v}\cdot \Delta t_{i-M} \Vert + b + k \cdot \Delta t_{i-M},\\\hspace{3.2cm} i=M+1,\cdots,2M,
	\end{matrix} 
	\right.
\end{align}
where $[\cdot]_{i}$ is the $i$-th element of a vector.

The LAS problem can be solved using an ML estimator. Recall that the measurement noises $\varepsilon_{d_i}$ and $\varepsilon_{\rho_i}$ are independent and follow Gaussian distributions. The ML estimator is equivalent to a WLS minimization problem as
\begin{equation} \label{eq:UVDMLminimizer}
	\hat{\boldsymbol{\theta}}_{}=\text{arg}\min\limits_{{\boldsymbol{\theta}}_{}}\left\Vert\boldsymbol{\tau} - \mathit{h}({\boldsymbol{\theta}}_{})\right\Vert_{\bm{W}}^2
	\text{,}
\end{equation}
where $\hat{\boldsymbol{\theta}}_{}$ is the estimator, and  $\bm{W}$ is a positive-definite diagonal weighting matrix given by
\begin{equation} \label{eq:weightmatall}
	\bm{W}=  \mathrm{diag}\left(\frac{1}{\sigma_{d_1}^2},\cdots,\frac{1}{\sigma_{d_M}^2},\frac{1}{\sigma_{\rho_1}^2},\cdots,\frac{1}{\sigma_{\rho_M}^2}\right) \text{,}
\end{equation}
and $\Vert \boldsymbol{x}\Vert _{\bm{W}}^2=\boldsymbol{x}^T\bm{W}\boldsymbol{x}$.

This ML based method using the sequential Doppler shift and TOA measurements can solve the LAS problem for a moving UD with clock drift in the TDBS. We name it by LAS-SDT method.

\subsubsection{Iterative WLS Algorithm for LAS-SDT} \label{ILSPM-UVD}
We develop an iterative WLS algorithm based on the commonly adopted Gauss-Newton method \cite{kaplan2005understanding,misra2006global}. We first conduct a Taylor series expansion at the estimate point of
$
\check{\boldsymbol{\theta}}_{}=\left[\check{\boldsymbol{p}}^T,\check{b},\check{\boldsymbol{v}}^T,\check{k}\right]^T \text{,}
$
where $\check{\boldsymbol{p}}$, $\check b$, $\check{\boldsymbol{v}}$, and $\check{k}$ are estimates for ${\boldsymbol{p}}$, $b$, $\boldsymbol{v}$, and $k$, respectively. We ignore the second and higher order terms. Then, \eqref{eq:allmeasurements} becomes
\begin{equation} \label{eq:taylor}
	\boldsymbol{\tau} = \mathit{h}(\check{\boldsymbol{\theta}}_{}) + \left(\frac{\partial \mathit{h}(\boldsymbol{\theta}_{})}{\partial \boldsymbol{\theta}_{}}|_{\boldsymbol{\theta}_{}=\check{\boldsymbol{\theta}}_{}}\right)\left(\boldsymbol{\theta}_{}-\check{\boldsymbol{\theta}}_{}\right)+\boldsymbol\varepsilon \text{.}
\end{equation}

We define the error vector:
$$\Delta\boldsymbol{\theta}_{} \triangleq \boldsymbol{\theta}_{}-\check{\boldsymbol{\theta}}_{} \text{,}
$$
and the residual vector:
\begin{equation}\label{eq:residual}
	\boldsymbol{r} \triangleq \boldsymbol{\tau} - \mathit{h}(\check{\boldsymbol{\theta}}_{})= \check{\bm{G}} \cdot \Delta\boldsymbol{\theta}_{}+\boldsymbol\varepsilon \text{,}
\end{equation}
where $\check{\bm{G}}=\frac{\partial \mathit{h}(\boldsymbol{\theta}_{})}{\partial \boldsymbol{\theta}_{}}|_{\boldsymbol{\theta}_{}=\check{\boldsymbol{\theta}}_{}}$ is the estimated version of the design matrix ${\bm{G}}$ with $\check{\boldsymbol{\theta}}$ plugged in, and
\begin{align} \label{eq:GUderivative}
	&[{\bm{G}}]_{i,:} =\nonumber\\
	&\left\{
	\begin{matrix*}[l]
		\left[\frac{\left[\partial h\right]_{i}}{\partial \boldsymbol{p}},0,\frac{\left[\partial h\right]_{i}}{\partial \boldsymbol{v}},1\right],i=1,\cdots,M,\\
		\left[-{\boldsymbol{e}}_{i-M}^T,1,-{\boldsymbol{e}}_{i-M}^T\Delta t_{i-M},\Delta t_{i-M} \right], i=M+1,\cdots,2M, \\
	\end{matrix*}
	\right.
\end{align}
\begin{align}\label{partialp}
	&\frac{\left[\partial h\right]_{i}}{\partial \boldsymbol{p}}=
	\frac{\boldsymbol{v}^T}{\Vert \boldsymbol{q}_i - {\boldsymbol{p}}-{\boldsymbol{v}} \Delta t_i\Vert}\nonumber\\&-
	\frac{\boldsymbol{v}^T(\boldsymbol{q}_i - {\boldsymbol{p}}-{\boldsymbol{v}} \Delta t_i)(\boldsymbol{q}_i - {\boldsymbol{p}}-{\boldsymbol{v}} \Delta t_i)^T}{\Vert \boldsymbol{q}_i - {\boldsymbol{p}}-{\boldsymbol{v}} \Delta t_i\Vert^3}, i=1,\cdots,M,
\end{align}
\begin{align}\label{partialv}
	&\frac{\left[\partial h\right]_{i}}{\partial \boldsymbol{v}}=
	\frac{2\boldsymbol{v}^T \Delta t_i+\boldsymbol{p}^T-\boldsymbol{q}_i^T}{\Vert \boldsymbol{q}_i - {\boldsymbol{p}}-{\boldsymbol{v}} \Delta t_i\Vert}\nonumber\\
	&-\frac{\Delta t_i \boldsymbol{v}^T(\boldsymbol{q}_i - {\boldsymbol{p}}-{\boldsymbol{v}} \Delta t_i)(\boldsymbol{q}_i - {\boldsymbol{p}}-{\boldsymbol{v}} \Delta t_i)^T}{\Vert \boldsymbol{q}_i - {\boldsymbol{p}}-{\boldsymbol{v}} \Delta t_i\Vert^3}, i=1,\cdots,M,
\end{align}
\begin{equation} \label{eq:LOS}
	{\boldsymbol{e}}_{i}=\frac{\boldsymbol{q}_i - {\boldsymbol{p}}-{\boldsymbol{v}} \Delta t_i}{\Vert \boldsymbol{q}_i - {\boldsymbol{p}}-{\boldsymbol{v}} \Delta t_i\Vert } \text{, } i=1,\cdots,M,
\end{equation}
with  $[\cdot]_{i,:}$ denoting the $i$-th row of a matrix, and ${\boldsymbol{e}}_{i}$ representing the unit line-of-sight (LOS) vector from the UD to AN \#$i$.

We estimate the error vector $\Delta\boldsymbol{\theta}$ in a WLS sense, and denote the estimated parameter error by $\Delta \check{\boldsymbol{\theta}}$. It is given by
\begin{equation} \label{eq:leastsquare}
	\Delta\check{\boldsymbol{\theta}}=(\check{\bm{G}}^T\bm{W}\check{\bm{G}})^{-1}\check{\bm{G}}^T\bm{W}\boldsymbol{r} \text{.}
\end{equation}

The estimated parameter vector is updated by
\begin{equation} \label{eq:thetaupdate}
	\check{\boldsymbol{\theta}} \leftarrow \check{\boldsymbol{\theta}} + \Delta \check{\boldsymbol{\theta}} \text{.}
\end{equation}

We substitute \eqref{eq:thetaupdate} into \eqref{eq:residual} and compute the estimated parameter $\check{\boldsymbol{\theta}}$ iteratively until convergence or a iteration count limit is reached. The iterative WLS algorithm for the LAS-SDT method is summarized in Algorithm 1.

The proposed iterative algorithm for the LAS-STD requires a proper initialization to guarantee the convergence to the correct solution. In real-world applications, we can utilize some prior knowledge such as a rough estimate or the known UD position from the previous estimation as the initialization. To the best of the authors’ knowledge, the proposed method is the first one to solve the LAS problem based on sequential Doppler shift and TOA measurements. In the future, we will investigate possible alternative solutions, such as the closed-form methods and semi-definite programming (SDP) methods.



\begin{algorithm}
	\caption{LAS-SDT}
	\label{al:unknownV}
	\begin{algorithmic}[1]
		\State Input: Doppler shift measurements $\boldsymbol{d}$ and TOA measurements $\boldsymbol{\rho}$, noise variance $\bm{\Sigma}_{d}$ and $\bm{\Sigma}_{\rho}$, ANs' positions $\boldsymbol{q}_i$, $i=1,\cdots,M$, initial parameter estimate $\check{\boldsymbol{\theta}}_{0}=[\check{\boldsymbol{p}}_{0}^T,\check b_0, \check{\boldsymbol{v}}_0^T,  \check k_0]^T$, maximum iterative count $iter$, and convergence threshold $thr$.
		\For {$s=1:iter$}
		\State Calculate unit LOS vector $\check{\boldsymbol{e}}_{Ui}$ based on (\ref{eq:LOS}), $i=1,\cdots,M$
		\State Compute residual vector  $\boldsymbol{r}$ using \eqref{eq:residual}
		\State Form design matrix $\check{\bm{G}}$ based on (\ref{eq:GUderivative})
		\State Calculate estimated parameter error vector $\Delta\check{\boldsymbol{\theta}}$ using (\ref{eq:leastsquare})
		\State Update parameter estimate $\check{\boldsymbol{\theta}}_{s} = \check{\boldsymbol{\theta}}_{s-1} + \Delta \check{\boldsymbol{\theta}}$
		\If {$\Vert [\Delta \check{\boldsymbol{\theta}}]_{1:N+1} \Vert<thr$}
		\State Exit \textbf{for} loop
		\EndIf
		\EndFor
		\State Output: $\check{\boldsymbol{\theta}}_{s}$
	\end{algorithmic}
\end{algorithm}

\subsection{Optimal Estimator with UD Velocity Aiding}
\subsubsection{LAS-SDT-v}\label{MLwithV}
The UD velocity during one broadcast round can be obtained by some other sensors such as an inertial measurement unit (IMU) or an optical flow sensor. With this aiding information of the UD velocity, we can use an ML estimator to solve the LAS problem. We name this method by LAS-SDT with velocity aiding or LAS-SDT-v for short.

In practice, the aiding velocity information, denoted by $\tilde{\bm{v}}$, may be subject to error, denoted by $\bm{\varepsilon}_v$. We model the error as a zero-mean Gaussian noise with a variance of $\bm{\Sigma}_v$, i.e., $\bm{\varepsilon}_{v} \sim \mathcal{N}(0,\bm{\Sigma}_{v})$.
We denote the parameters to be estimated by $\boldsymbol{\theta}_{v}$, as given by 
$
\boldsymbol{\theta}_{v}=\bm{\theta}
$.

The relation between all the measurements and the parameter $\boldsymbol{\theta}_v$ reads
\begin{equation} \label{eq:MLvectorpr}
	\bm{z}_v = y_v(\boldsymbol{\theta}_{v}) + \left[\begin{matrix}
	\boldsymbol{\varepsilon}\\
	\bm{\varepsilon}_v
\end{matrix}\right] \text{,}
\end{equation}
where 
\begin{align}
	\bm{z}_v=\left[\begin{matrix}
		\boldsymbol{\tau}\\
		\tilde{\bm{v}}
	\end{matrix}\right], \; y_v(\boldsymbol{\theta}_{v})=\left[\begin{matrix}
	h(\boldsymbol{\theta}_{v})\\
	\bm{v}
\end{matrix}
\right],
\end{align}
and the function $h(\cdot)$ has the same form as \eqref{eq:funtheta}.

The parameter $\boldsymbol{\theta}_v$ is estimated by solving the WLS minimization problem as
\begin{equation} \label{eq:MLminimizer}
	\hat{\boldsymbol{\theta}}_{v}=\text{arg}\min\limits_{{\boldsymbol{\theta}}_{v}} \left\Vert\bm{z}_v - y_v(\boldsymbol{\theta}_{v}) \right\Vert_{\bm{W}_v}^2
	\text{,}
\end{equation}
where $\hat{\boldsymbol{\theta}}_{v}$ is the estimator, and
$$
\bm{W}_v=\left[\begin{matrix}
	\bm{W} & \\
	& \bm{\Sigma}_v^{-1}
\end{matrix}\right].
$$

%

\subsubsection{Iterative WLS Algorithm for LAS-SDT-v}
The iterative algorithm is similar to the LAS-SDT in Algorithm \ref{al:unknownV}. The differences are the estimated unit LOS vector $\check {\boldsymbol{e}}_{v_i}$, estimated design matrix $\check {\bm{G}}_{v}$, and estimated error vector $\Delta \check{\boldsymbol{\theta}}_v$, as given by
\begin{equation} \label{eq:LOSK}
	\check{\boldsymbol{e}}_{v_i}=\frac{\boldsymbol{q}_i - \check{\boldsymbol{p}}-\tilde{\boldsymbol{v}} \Delta t_i}{\Vert \boldsymbol{q}_i - \check{\boldsymbol{p}}-\tilde{\boldsymbol{v}} \Delta t_i\Vert } \text{,}
\end{equation}
\begin{align} \label{eq:GKderivative}
	[\check{\bm{G}}_v]_{i,:} =\left[\begin{matrix}
		\check{\bm{G}}\\
		\begin{matrix}
			\bm{O}_{N\times(N+1)}&\bm{I}_N&\bm{0}_{N}
		\end{matrix}
	\end{matrix}\right],
\end{align}
and
\begin{equation} \label{eq:leastsquareK}
	\Delta\check{\boldsymbol{\theta}}_v=(\check{\bm{G}}_v^T\bm{W}_v\check{\bm{G}}_v)^{-1}\check{\bm{G}}_v^T\bm{W}_v\boldsymbol{r}_v \text{.}
\end{equation}

The iterative process of the LAS-SDT-v is summarized in Algorithm \ref{al:knownV}.

\begin{algorithm}
	\caption{LAS-SDT-v}
	\label{al:knownV}
	\begin{algorithmic}[1]
		\State Input: Doppler shift measurements $\boldsymbol{d}$ and TOA measurements $\boldsymbol{\rho}$, noise variance  $\bm{\Sigma}_{d}$ and $\bm{\Sigma}_{\rho}$, ANs' positions $\boldsymbol{q}_i$, $i=1,\cdots,M$, aiding UD velocity $\boldsymbol{v}$, velocity error variance $\bm{\Sigma}_v$, initial parameter estimate $\check{\boldsymbol{\theta}}_{v_0}=[\check{\boldsymbol{p}}_{0}^T, \check b_0, \check{\bm{v}}_0^T,\check{k}_0]^T$, maximum iterative count $iter$, and convergence threshold $thr$.
		\For {$s=1:iter$}
		\State Calculate unit LOS vector $\check{\boldsymbol{e}}_{v_i}$ based on (\ref{eq:LOSK}), $i=1,\cdots,M$
		\State Compute residual vector $\boldsymbol{r}_v=\bm{z}_v - y_v(\check{\boldsymbol{\theta}}_{v_{s-1}})$
		\State Form design matrix $\check{\bm{G}}_v$ based on (\ref{eq:GKderivative})
		\State Calculate estimated error vector $\Delta \check{\boldsymbol{\theta}}_v$ using (\ref{eq:leastsquareK})
		\State Update parameter estimate $\check{\boldsymbol{\theta}}_{v_s} = \check{\boldsymbol{\theta}}_{v_{s-1}} + \Delta \check{\boldsymbol{\theta}}_v$
		\If {$\Vert [\Delta \check{\boldsymbol{\theta}}_v]_{1:N+1} \Vert<thr$}
		\State Exit \textbf{for} loop
		\EndIf
		\EndFor
		\State Output: $\check{\boldsymbol{\theta}}_{v_s}$
	\end{algorithmic}
\end{algorithm}

\subsection{Optimal Estimator with UD Clock Drift Aiding}
\subsubsection{LAS-SDT-k}\label{MLwithome}
Current oscillators, even the consumer-level products, usually have a good frequency stability. This enables us to obtain the oscillator frequency offset or the clock drift through multiple UD-AN communications when the UD is stationary \cite{zhao2021parn}. In this case with UD clock drift aiding, we develop an ML-based LAS method, namely LAS-SDT-k.

We denote the aiding clock drift by $\tilde{{k}}$, which is subject to error, denoted by $\bm{\varepsilon}_k$. We model the error as a zero-mean Gaussian noise with a variance of $\sigma_k^2$, and $\bm{\varepsilon}_{k} \sim \mathcal{N}(0,\sigma_k^2)$. The parameter to be estimated is denoted by $\boldsymbol{\theta}_{k}$, and $
\boldsymbol{\theta}_{k}=\bm{\theta}
$.


The parameter $\boldsymbol{\theta}_{k}$ is estimated by solving the WLS minimization problem as
\begin{equation} \label{eq:OmeMLminimizer}
	\hat{\boldsymbol{\theta}}_{k}=\text{arg}\min\limits_{{\boldsymbol{\theta}}_{k}} \left\Vert\boldsymbol{z}_k - \mathit{y}_k({\boldsymbol{\theta}}_{k})\right\Vert_{\bm{W}_k}^2
	\text{,}
\end{equation}
where $\hat{\boldsymbol{\theta}}_{k}$ is the estimator,
\begin{align}
	\bm{z}_k=\left[\begin{matrix}
			\boldsymbol{\tau}\\
			\tilde{k}
		\end{matrix}\right], \; {y}_k(\boldsymbol{\theta}_{k})=\left[\begin{matrix}
			h(\boldsymbol{\theta}_{k})\\
			k
		\end{matrix}
		\right],
	\bm{W}_k=\left[\begin{matrix}
		\bm{W} & \\
		& \sigma_k^{-2}
	\end{matrix}\right],
\end{align}
and $h(\cdot)$ has the same form as given by \eqref{eq:funtheta}.

%

\subsubsection{Iterative WLS Algorithm for LAS-SDT-k}
The iterative algorithm is similar Algorithm \ref{al:unknownV}. However, the estimated design matrix $\check {\bm{G}}_{k}$, and estimated error vector $\Delta \check{\boldsymbol{\theta}}_{k}$ have different forms, as given by
\begin{align} \label{eq:Goderivative}
	[\check{\bm{G}}_{k}]_{i,:} =\begin{bmatrix}
		\check{\bm{G}}\\
		\begin{matrix}
			\bm{0}_{2N+1}^T&1
		\end{matrix}
	\end{bmatrix},
\end{align}
and
\begin{equation} \label{eq:leastsquareO}
	\Delta\check{\boldsymbol{\theta}}_{k}=(\check{\bm{G}}_{k}^T\bm{W}_k\check{\bm{G}}_{k})^{-1}\check{\bm{G}}_{k}^T\bm{W}_k\boldsymbol{r}_{k} \text{.}
\end{equation}

Algorithm \ref{al:knownome} is a summary of the iterative approach of the LAS-SDT-k.

\begin{algorithm}
	\caption{LAS-SDT-k}
	\label{al:knownome}
	\begin{algorithmic}[1]
		\State Input: Doppler measurements $\boldsymbol{d}$ and TOA measurements $\boldsymbol{\rho}$, noise variance $\bm{\Sigma}_{d}$ and $\bm{\Sigma}_{\rho}$, ANs' positions $\boldsymbol{q}_i$, $i=1,\cdots,M$, aiding UD clock drift $k$, initial parameter estimate $\check{\boldsymbol{\theta}}_{k_0}=[\check{\boldsymbol{p}}_{0}^T,\check b_0, \check{\boldsymbol{v}}_0^T,\check{k}_0]^T$, clock drift error variance $\bm{\Sigma}_k$, maximum iterative count $iter$, and convergence threshold $thr$.
		\For {$s=1:iter$}
		\State Calculate unit LOS vector $\check{\boldsymbol{e}}_{i}$ based on (\ref{eq:LOS}), $i=1,\cdots,M$
		\State Compute residual vector $\boldsymbol{r}_{k}=\boldsymbol{z}_k - \mathit{y}_k(\check{\boldsymbol{\theta}}_{k_{s-1}})$
		\State Form design matrix $\check{\bm{G}}_{k}$ based on (\ref{eq:Goderivative})
		\State Calculate estimated error vector $\Delta \check{\boldsymbol{\theta}}_{k}$ using (\ref{eq:leastsquareO})
		\State Update parameter estimate $\check{\boldsymbol{\theta}}_{{k}_s} = \check{\boldsymbol{\theta}}_{{k}_{s-1}} + \Delta \check{\boldsymbol{\theta}}_{k}$
		\If {$\Vert [\Delta \check{\boldsymbol{\theta}}_{k}]_{1:N+1} \Vert<thr$}
		\State Exit \textbf{for} loop
		\EndIf
		\EndFor
		\State Output: $\check{\boldsymbol{\theta}}_{{k}_s}$
	\end{algorithmic}
\end{algorithm}

\section{LAS Performance Analysis} \label{locanalysis}

\subsection{LAS-SDT Estimation Error Analysis} \label{Errorana}
\subsubsection{Estimation Error} \label{errorKVD}
We denote the estimation bias of the LAS-SDT by $\boldsymbol{\mu}$, which has $2N+2$ elements. Note that an ML estimator is asymptotically unbiased \cite{kay1993fundamentals}. Therefore, we have
\begin{align}
	\boldsymbol{\mu}=\boldsymbol{0}.
\end{align}



The estimation error variance, denoted by $\bm{Q}$, is 
\begin{align}
	\bm{Q}&=\mathbb{E}\left[\left(\Delta\boldsymbol{\theta}-\mathbb{E}[\Delta\boldsymbol{\theta}]\right)\left(\Delta\boldsymbol{\theta}-\mathbb{E}[\Delta\boldsymbol{\theta}]\right)^T\right]\nonumber\\ &=(\bm{G}^T\bm{W}\bm{G})^{-1}.
\end{align}

The root mean square error (RMSE) is thereby
\begin{align}
	\text{RMSE} &=\sqrt{\Vert\boldsymbol{\mu}\Vert^2+\mathrm{tr}(\bm{Q})}=\sqrt{\mathrm{tr}(\bm{Q})} \text{,}
\end{align}
where $\mathrm{tr}(\cdot)$ is trace of a matrix.

We notice that the weighting matrix $\bm{W}$ is comprised of the reciprocals of the measurement noise variances. Then the estimation error variance $\bm{Q}$ is growing quadratically with the increasing measurement noise $\sigma_{d_i}$ and $\sigma_{\rho_i}$. Therefore, the RMSE grows linearly when the measurement noise $\sigma_{d_i}$ and $\sigma_{\rho_i}$ increase.

\subsubsection{CRLB Derivation for LAS-SDT}\label{CRLBderive}
With $M$ Doppler shifts and $M$ TOA measurements at the UD, we have the likelihood function, denoted by $f(\boldsymbol{\tau}|\boldsymbol{\theta})$, as
\begin{equation} \label{eq:Ulikelihood}
	f(\boldsymbol{\tau}|\boldsymbol{\theta}) = \frac{\exp\left(-\frac{1}{2} \Vert\boldsymbol{\tau} - \mathit{h}({\boldsymbol{\theta}})\Vert_{\bm{W}}^2\right)}{(2\pi)^{M} |\bm{W}^{-1}|^{\frac{1}{2}}} \text{.}
\end{equation}



The Fisher information matrix (FIM) denoted by $\mathcal{F}$ is 
\begin{equation} \label{eq:UFIMvsG}
	\mathcal{F}=-\mathbb{E}\left[\frac{\partial^2 \ln f(\boldsymbol{\rho}|\boldsymbol{\theta})}{\partial\boldsymbol{\theta} \partial\boldsymbol{\theta}^T}\right]= \bm{G}^T\bm{W}\bm{G} \text{,}
\end{equation}
in which $\mathbb{E}[\cdot]$ is the expectation operator, and $\bm{G}_{}$ is given by (\ref{eq:GUderivative}).


The CRLB for the $i$-th element in $\boldsymbol{\theta}$ is expressed by 
\begin{equation} \label{eq:UCRLBFisher}
	\mathsf{CRLB}([\boldsymbol{\theta}]_i)=[\mathcal{F}^{-1}]_{i,i} \text{,}
\end{equation}
where  $[\cdot]_{i}$ represents the $i$-th element of a vector, and $[\cdot]_{i,i}$ represents the diagonal element of a matrix at the $i$-th column and the $i$-th row.

\hypertarget{R1}{\textbf{Remark 1}}: The estimation accuracy of the LAS-SDT is higher than that of the conventional sequential TOA localization method such as the LSPM-UVD \cite{zhao2020optloc}. Intuitively, the LAS-SDT utilizes Doppler shifts in addition to TOA measurements and thus has a better performance. This is also proven mathematically in Appendix \ref{Appendix1}.

\subsection{LAS-SDT-v Estimation Error Analysis}
\subsubsection{Estimation Error and CRLB}\label{knownV1}
We denote the estimation bias of the LAS-SDT-v by $\boldsymbol{\mu}_v$. Similar to the bias of LAS-SDT, we have
\begin{align}
	\boldsymbol{\mu}_v=\boldsymbol{0}.
\end{align}

The estimation error variance, denoted by $\bm{Q}_v$ is
\begin{align}
	\bm{Q}_v
	=(\bm{G}_v^T\bm{W}_v\bm{G}_v)^{-1}.
\end{align}
where
\begin{align} \label{eq:Gvtruederivative}
	[{\bm{G}}_v]_{i,:}= \left[\begin{matrix}
			{\bm{G}}\\
			\begin{matrix}
				\bm{O}_{N\times(N+1)}&\bm{I}_N&\bm{0}_{N}
			\end{matrix}
		\end{matrix}\right].
\end{align}

The RMSE is
\begin{align}
	\text{RMSE}_v &=\sqrt{\Vert\boldsymbol{\mu}_v\Vert^2+\mathrm{tr}(\bm{Q}_v)}=\sqrt{\mathrm{tr}(\bm{Q}_v)} \text{.}
\end{align}

Similar to the CRLB derivation for the LAS-SDT in Section \ref{CRLBderive}, the CRLB for LAS-SDT-v, denoted by $\mathsf{CRLB}_v$ is
\begin{equation} \label{eq:KVCRLBFisher}
	\mathsf{CRLB}_v\left([\boldsymbol{\theta}_v]_i\right)=[\mathcal{F}_v^{-1}]_{i,i}=\left[(\bm{G}_v^T\bm{W}_v\bm{G}_v)^{-1}\right]_{i,i} \text{,}
\end{equation}
where $\mathcal{F}_v$ is the FIM for LAS-SDT-v.

\hypertarget{R1}{\textbf{Remark 2}}: Compared with the LAS-SDT, the aiding velocity $\bar{\boldsymbol{v}}$ in the LAS-SDT-v is treated as measurements with extra information. Therefore, with more measurements, we can intuitively know that there is a performance gain in the LAS-SDT-v, i.e., the estimation error of the LAS-SDT-v is smaller than that of the LAS-SDT. When the velocity error approaches infinite, there will be little information in the aiding velocity and the estimation error of the LAS-SDT-v will approach that of the LAS-SDT. It is proven mathematically in Appendix \ref{Appendix2}.

\subsubsection{Estimation Error Caused by Deviated Velocity Information} \label{deviatedV}
In real-world applications, the UD velocity measured by a sensor may not be accurate enough, resulting in deviated velocity information from the true value. This will cause errors in the LAS estimation.

The deviated aiding UD velocity is denoted by $\bar{\boldsymbol{v}}$.
The deviated velocity-caused error vector, denoted by $\bar{\boldsymbol{r}}_v$, is given by
\begin{align}\label{eq:resdV}
	\bar{\boldsymbol{r}}_{v} =
	\begin{bmatrix}
		\bm{0}_{2M}\\
		\bm{v}-\bar{\bm{v}}
	\end{bmatrix}
	\text{.}
\end{align}

Then, the estimation bias denoted by $\bar{\boldsymbol{\mu}}_{v}$ is 
\begin{equation}\label{eq:dPvsdV}
	\bar{\boldsymbol{\mu}}_{v}=({\bm{G}}_v^T\bm{W}_{v}{\bm{G}}_v)^{-1}{\bm{G}}_v^T\bm{W}_{v}\bar{\boldsymbol{r}}_v \text{,}
\end{equation}


The estimation variance, denoted by $\bar{\bm{Q}}_v$, and the RMSE, denoted by $\overline{\text{RMSE}}_v$,
is given by
\begin{equation} \label{eq:posvardV}
	\bar{\bm{Q}}_v=({\bm{G}}_v^T\bm{W}_{v}{\bm{G}}_v)^{-1} \text{,}
\end{equation}
and
\begin{equation} \label{eq:RMSEdV}
	\overline{\text{RMSE}}_v  =\sqrt{\Vert\bar{\boldsymbol{\mu}}_v\Vert^2+\mathrm{tr}(\bar{\bm{Q}}_v)} \text{,}
\end{equation}
respectively.

\hypertarget{R3}{\textbf{Remark 3}}: The estimation bias $\bar{\boldsymbol{\mu}}_v$ can be expressed as an increasing function of the deviation from the true velocity, denoted by $\Delta \boldsymbol{v}=\boldsymbol{v}-\bar{\boldsymbol{v}}$. It is shown in Appendix \ref{Appendix3}.

\subsection{LAS-SDT-k Estimation Error Analysis} \label{knownO1}
\subsubsection{Estimation Error and CRLB}
We denote the estimation bias of the LAS-SDT-k by $\boldsymbol{\mu}_{k}$ and have
\begin{align}
	\boldsymbol{\mu}_{k}=\boldsymbol{0}.
\end{align}

The estimation error variance, denoted by $\bm{Q}_{k}$ is
\begin{align}
	\bm{Q}_{k}
	=(\bm{G}_{k}^T\bm{W}_k\bm{G}_{k})^{-1}.
\end{align}
where
\begin{align} \label{eq:Gometruederivative}
	[{\bm{G}}_{k}]_{i,:} =\begin{bmatrix}
			{\bm{G}}\\
			\begin{matrix}
				\bm{0}_{2N+1}^T&1
			\end{matrix}
	\end{bmatrix}.
\end{align}

The RMSE is
\begin{align}
	\text{RMSE}_{k} &=\sqrt{\Vert\boldsymbol{\mu}_{k}\Vert^2+\mathrm{tr}(\bm{Q}_{k})}=\sqrt{\mathrm{tr}(\bm{Q}_{k})} \text{.}
\end{align}

The CRLB for LAS-SDT-k, denoted by $\mathsf{CRLB}_{k}$ is
\begin{equation} \label{eq:KomeCRLBFisher}
	\mathsf{CRLB}_{k}\left([\boldsymbol{\theta}_{k}]_i\right)=\left[(\bm{G}_{k}^T\bm{W}_k\bm{G}_{k})^{-1}\right]_{i,i} \text{.}
\end{equation}

Similar to LAS-SDT-v, the estimation errors of LAS-SDT-k are smaller than that of the LAS-SDT.
\subsubsection{Estimation Error Caused by Deviated Clock Drift Information}
The UD clock drift is determined by the oscillator frequency, which may vary with time and temperature. The LAS estimation accuracy will degrade if the aiding clock drift deviates from its true value.

We denote the aiding clock drift by $\bar{k}$, and the deviation by $\Delta k = k- \bar {k}$. Then the error vector, denoted by $\bar{\boldsymbol{r}}_{k}$, is
\begin{align}\label{eq:resome}
	\bar{\boldsymbol{r}}_{k}=
	\begin{bmatrix}
		\bm{0}_{2M}\\
		\Delta k 
	\end{bmatrix}.
\end{align}

The estimation bias, denoted by $\bar{\boldsymbol{\mu}}_{k}$, is
\begin{equation}\label{eq:dPvsdome}
	\bar{\boldsymbol{\mu}}_{k}=({\bm{G}}_{k}^T\bm{W}_k{\bm{G}}_{k})^{-1}{\bm{G}}_{k}^T\bm{W}_{}\bar{\boldsymbol{r}}_{k} \text{,}
\end{equation}
the estimation variance is
\begin{equation} \label{eq:posvardome}
	\bar{\bm{Q}}_{k}=({\bm{G}}_{k}^T\bm{W}_k{\bm{G}}_{k})^{-1} \text{,}
\end{equation}
and the RMSE is
\begin{equation} \label{eq:RMSEdome}
	\overline{\text{RMSE}}_{k}  =\sqrt{\Vert\bar{\boldsymbol{\mu}}_{k}\Vert^2+\mathrm{tr}(\bar{\bm{Q}}_{k})} \text{.}
\end{equation}

We can see from \eqref{eq:resome} and \eqref{eq:dPvsdome} that the estimation bias grows with the clock drift deviation. This can be proven similarly as Remark 3.

\section{Numerical Simulation}
\label{simulation}
We conduct numerical simulations to assess the LAS performance of the proposed LAS-SDT method. We use the CRLB as the benchmark to evaluate the estimation accuracy. In all the simulations, we compute the RMSE of the estimated parameters. We take the position result as an example, and have the RMSE as given by
\begin{align} \label{eq:RMSEdef}
	\text{RMSE}&=\sqrt{\frac{1}{K_s}\sum_{1}^{K_s}\Vert\boldsymbol{p}-\hat{\boldsymbol{p}}\Vert^2}
\end{align}
where $K_s$ is the number of simulation runs and $\hat{\boldsymbol{p}}$ is the localization result from the proposed method under test in each simulation.

\subsection{Simulation Settings}
\begin{figure}
	\centering
	\includegraphics[width=0.99\linewidth]{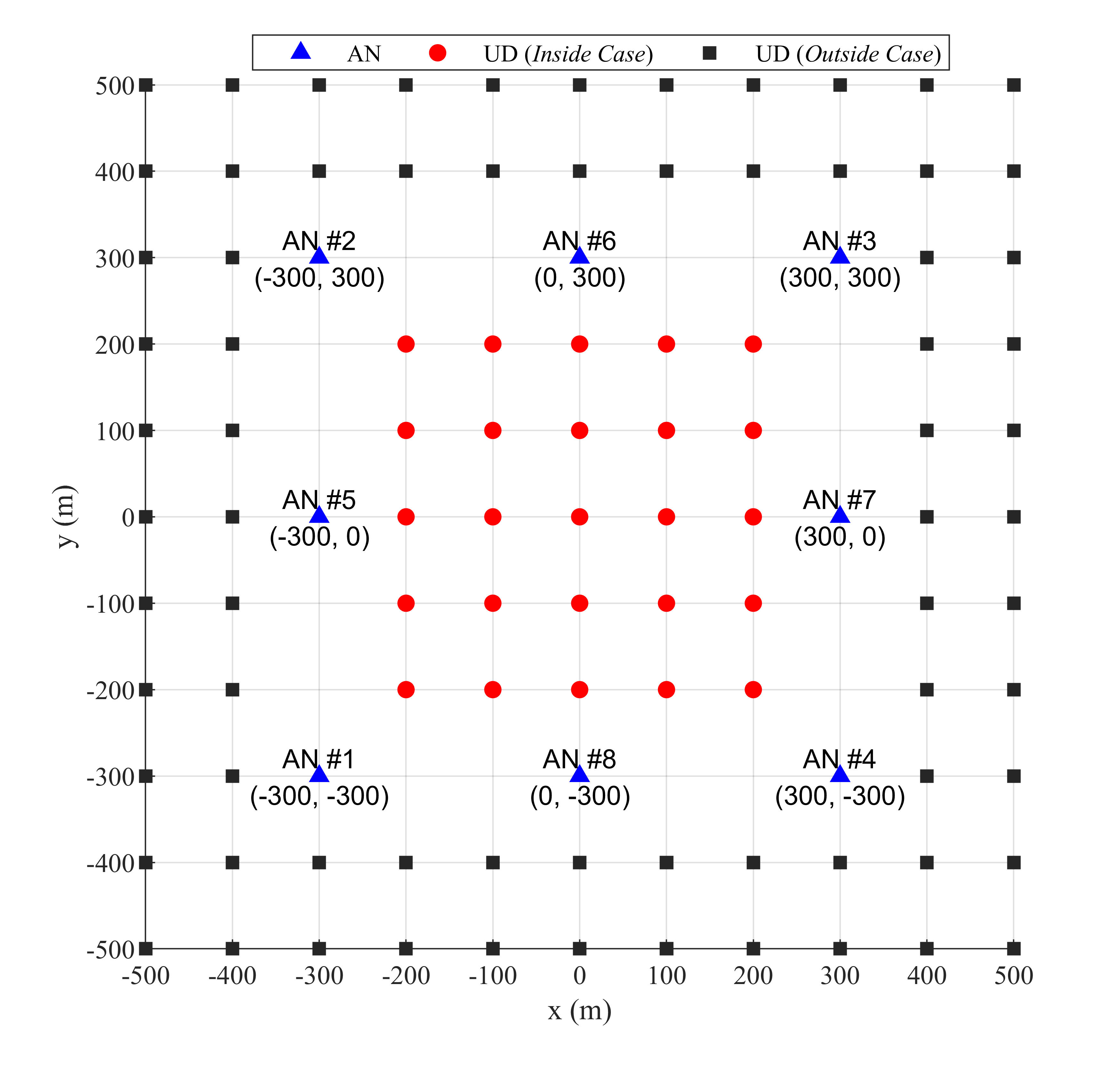}
	\caption{AN and UD placement in simulation scene. ANs are located at the vertices and the middle of the edges, and UD is randomly placed at the red dots (\textit{Inside Case}) or at the black squares (\textit{Outside Case}). }
	\label{fig:simulationsetting}
\end{figure}

We create a 2D simulation scene with 8 ANs on the corners and middle points on edges of a 600 m $\times$ 600 m square area as shown in Fig. \ref{fig:simulationsetting}. The UD locations have two cases, i.e., the \textit{Inside Case} and the \textit{Outside Case}. In the \textit{Inside Case}, the UD locates randomly at the red dots inside the convex hull formed by the ANs as given in Fig. \ref{fig:simulationsetting}. And for the \textit{Outside Case}, the UD locations are the black squares outside the AN convex hull as given in the figure. The UD speed $\Vert\boldsymbol{v}\Vert$ is randomly drawn from a uniform distribution $\mathcal{U}(0,50)$ m/s. The direction of the UD velocity is drawn from $\mathcal{U}(0,2\pi)$. The initial UD clock offset and drift are set randomly at the start of each simulation run. The clock offset $b$ is drawn from the uniform distribution $\mathcal{U}(-1,1)$ s. The clock drift $k$ is selected from $\mathcal{U}(-20,20)$ parts per million (ppm). The interval between successive Doppler shift or TOA measurements is set to 50 ms. 
We set the standard deviation (STD) of the TOA measurement noise $\sigma_{\rho}$ to vary from 0.1 m to 10 m with 5 steps. We set the Doppler shift measurement noise $\sigma_{d}$ to $5\sigma_{\rho}$ in m/s, which is at the same level of \cite{jia2020localization}. At every noise step, we conduct 5,000 Monte-Carlo simulations.

For the proposed iterative algorithms, the maximum iteration time $iter$ is set to be 10 and the convergence threshold $thr=10^{-2}$ m. The initial position $\check{\boldsymbol{p}}_0$ is set to a random point on the circumference of a 60-m radius circle centered at the true position, representing an inaccurate initial guess. The other initial parameters are $\check{b}_0 = \rho_1$, $\check{k}_0=0$ and $\check{\boldsymbol{v}}_0=0$.

\subsection{LAS-SDT Estimation Performance} \label{simLASSTOAD}


\subsubsection{Inside Case}
\begin{figure}
	\centering
	\includegraphics[width=0.99\linewidth]{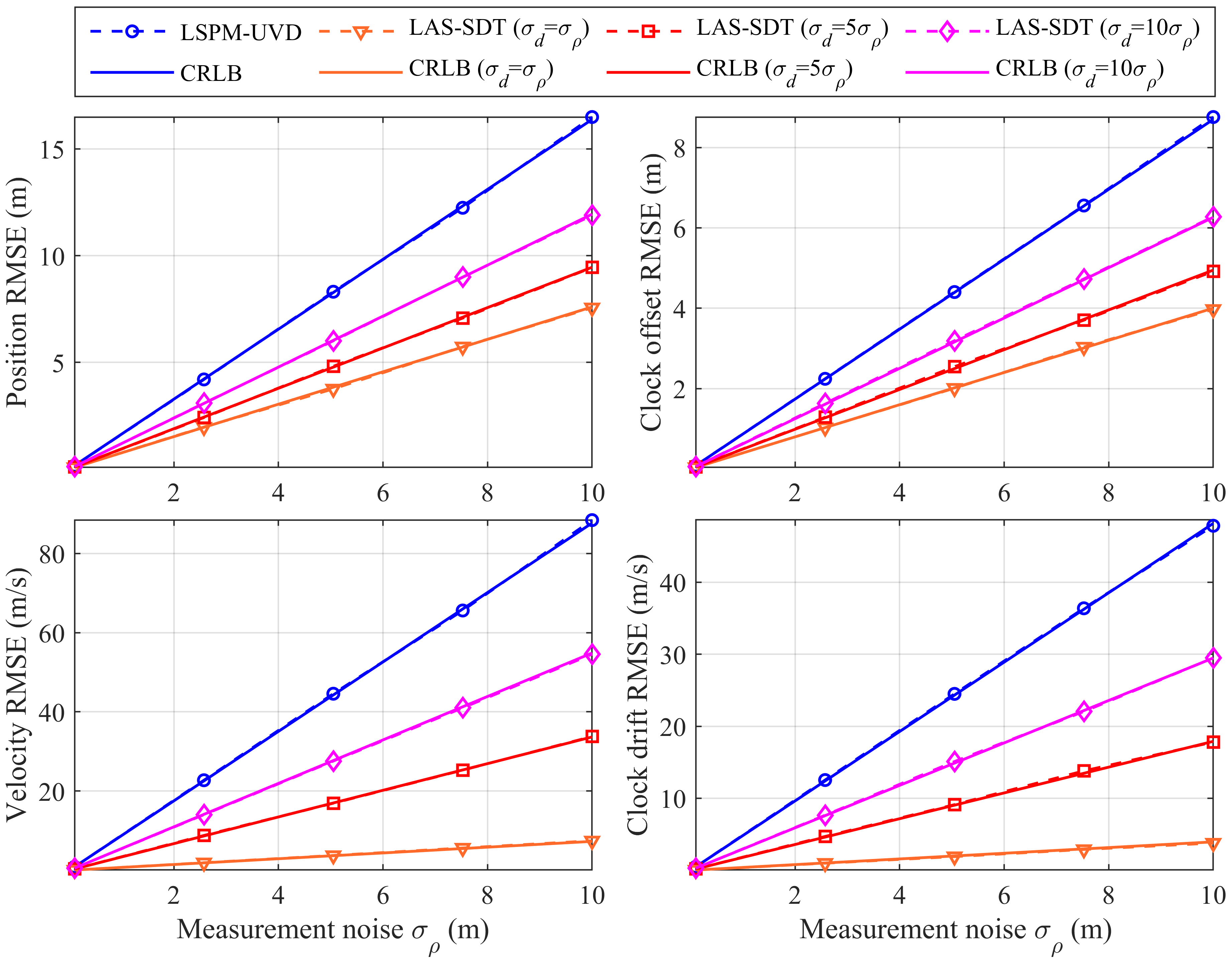} 
	\caption{Estimation error vs. measurement noise (\textit{Inside Case}). The estimation accuracy of the proposed LAS-SDT method reaches the CRLB. Compared with the conventional LSPM-UVD, the LAS-SDT has a significantly higher estimation accuracy due to the utilization of the Doppler shift measurements. With larger Doppler shift measurement noise, the LAS error increases.
	}
	\label{fig:RMSEvsnoise}
\end{figure}

In this simulation case, the UD is randomly placed at the red dots as shown in Fig. \ref{fig:simulationsetting}. The estimation results from the proposed LAS-SDT are shown in Fig. \ref{fig:RMSEvsnoise}. We also depict the results of the conventional method using sequential TOA measurements only, i.e., LSPM-UVD \cite{zhao2020optloc}, for comparison. Furthermore, the results with different Doppler shift measurement noises, i.e., $\sigma_{d}=\sigma_{\rho}$ and $\sigma_{d}=10\sigma_{\rho}$, are also shown in the figure for comparison. We can see that the estimation errors of the new LAS-SDT grow linearly with increasing measurement noise $\sigma_{\rho}$ and reach the CRLB. By utilizing the sequential Doppler shift measurements, all the estimation accuracies of the LAS-SDT for position, clock offset, velocity and clock drift are superior to those of the conventional LSPM-UVD. For example, the position and clock offset errors of the LAS-SDT are about 50\% smaller than that of the LSPM-UVD when $\sigma_{d}=5\sigma_{\rho}$. We also note that with larger Doppler shift measurement noise, the LAS error of the LAS-SDT will increase and approach that of the conventional LSPM-UVD. This is consistent with the theoretical analysis presented in Section \ref{Errorana}.


In order to investigate how the initialization error affects the convergence speed and estimation results of the iterative LAS-SDT, we set the initial position with different distances to the true position. The TOA measurement noise is fixed at $\sigma_{\rho}=10$ m and the Doppler shift measurement noise is set to $\sigma_d=5\sigma_{\rho}$. We record the average number of iterations under different initial position error conditions. The number of iterations and the localization RMSEs of the LAS-SDT method in the \textit{Inside Case} are shown in Table \ref{table_RMSE}.
	
As can be seen from Table \ref{table_RMSE}, when the initial position error is smaller than 100 m, the localization error of the new LAS-SDT method reaches the CRLB in the simulation scene. When the initial position error grows, the new method may not give an accurate solution and the estimation RMSE deviates from the CRLB. Furthermore, the average number of iterations shows that the LAS-SDT method will use more iterations to obtain the solution when the initialization error is larger.

\begin{table}[!t]
	\centering
	\begin{threeparttable}
		\caption{Number of Iterations and Localization RMSE for LAS-SDT with Different Initial Position Errors}
		\label{table_RMSE}
		\centering
		\begin{tabular}{c r r r r r  }
				\toprule
				 Initial Position Error (m) &60 & 100 & 200 & 300\\
				\hline
				Avg. number of iterations &3.83 &4.03 &4.48 &5.04\\
				RMSE (m) &9.48 &9.49 &\textbf{11.26} &\textbf{12.06}\\
				{CRLB (m)} &\multicolumn{4}{c}{9.45}
				\\				\bottomrule
		\end{tabular}
		\begin{tablenotes}[para,flushleft]
			Note: When the initial position error grows, such as 200 m and 300 m, the estimated localization RMSE will become significantly large and deviate from the CRLB, and the algorithm will spend more iterations to reach a solution.
		\end{tablenotes}
	\end{threeparttable}
\end{table}

\subsubsection{Outside Case}
\begin{figure}
	\centering
	\includegraphics[width=0.99\linewidth]{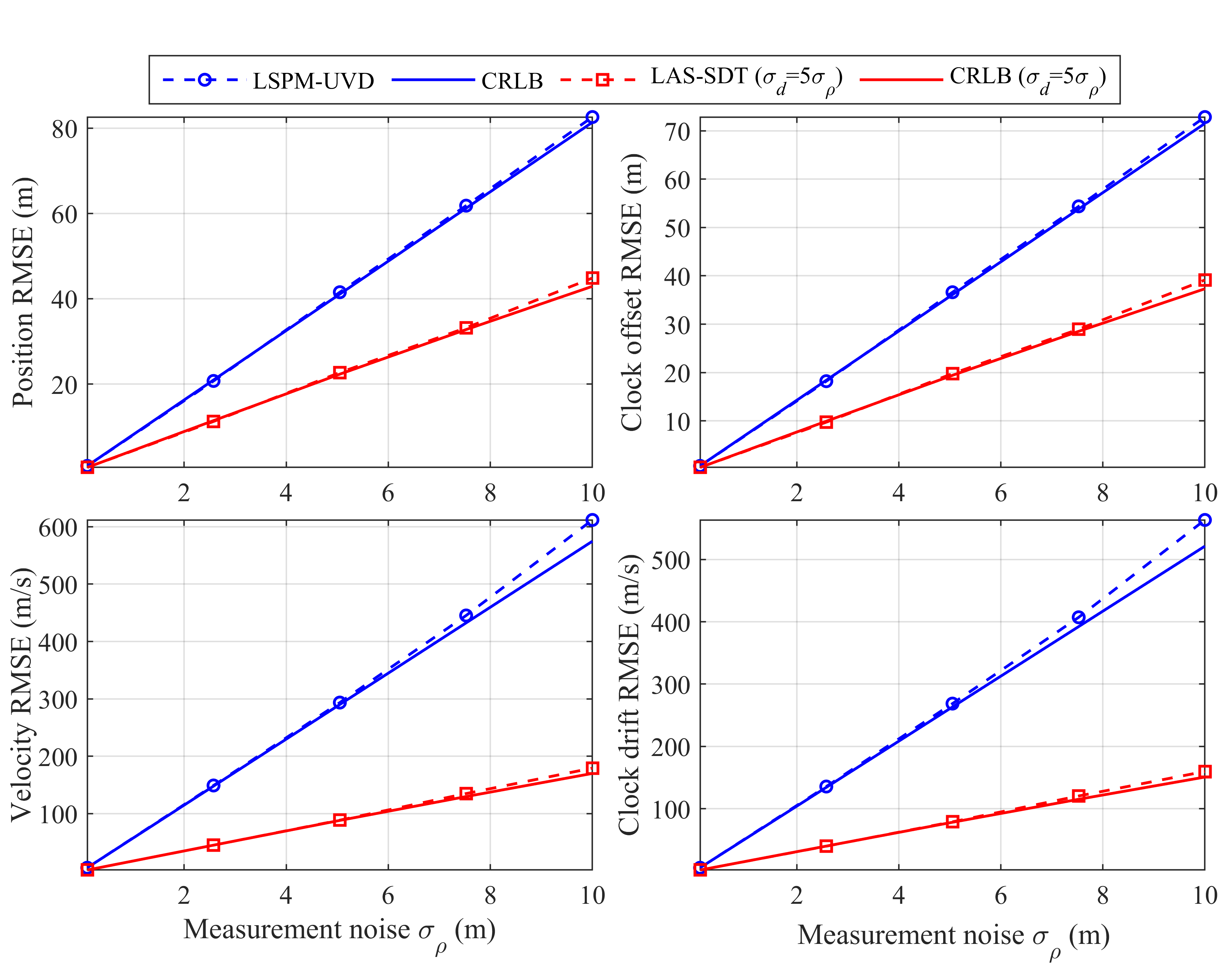} 
	\caption{Estimation error vs. measurement noise (\textit{Outside Case}). The estimation accuracy of the proposed LAS-SDT method reaches the CRLB in the small noise range. When the measurement noise becomes larger, the estimation errors tend to deviate from the CRLB. The LAS-SDT has significantly smaller estimation error compared with the conventional LSPM-UVD.
	}
	\label{fig:RMSEvsnoiseOutside}
\end{figure}

We investigate the performance of the new LAS-SDT in another practical case, i.e., the \textit{Outside Case}. The UD position is randomly selected from the black squares in Fig. \ref{fig:simulationsetting}. The estimation errors for position, velocity, clock offset and clock drift are shown in Fig. \ref{fig:RMSEvsnoiseOutside}. We also depict the results of the conventional LSPM-UVD in the same figure for comparison. 

We can see that Fig. \ref{fig:RMSEvsnoiseOutside} shows similar patterns as Fig. \ref{fig:RMSEvsnoise}. For example, the estimation accuracy of the new LAS-SDT method is significantly better than that of the conventional LSPM-UVD, showing the performance improvement using the Doppler shift measurements. We shall notice two differences between Fig. \ref{fig:RMSEvsnoiseOutside} and Fig. \ref{fig:RMSEvsnoise}, i.e., i) the estimation RMSEs of both the new LAS-SDT method and the conventional LSPM-UVD in the \textit{Outside Case} are larger than those in the \textit{Inside Case}, and ii) the estimation errors in the \textit{Outside Case} tend to deviate from the CRLBs when the measurement noise becomes larger. They are caused by the worse relative geometry between the UD and ANs in the \textit{Outside Case}, which results in a larger dilution of precision (DOP) \cite{kaplan2005understanding} and amplifies the estimation error.

\subsection{LAS Performance with UD Velocity Aiding (LAS-SDT-v)} \label{knownV}
\subsubsection{LAS with Aiding Velocity Subject to Random Error}
We set the aiding UD velocity to the true value added with zero-mean Gaussian random noise to investigate the estimation performance of the LAS-SDT-v with erroneous known velocity. The UD positions are randomly selected from the \textit{Inside Case}. The STD for the aiding velocity error is set to $\sigma_{v}=0.1\sigma_{\rho}$ in m/s. The other settings are the same as Section \ref{simLASSTOAD}.

\begin{figure}
	\centering
	\includegraphics[width=0.99\linewidth]{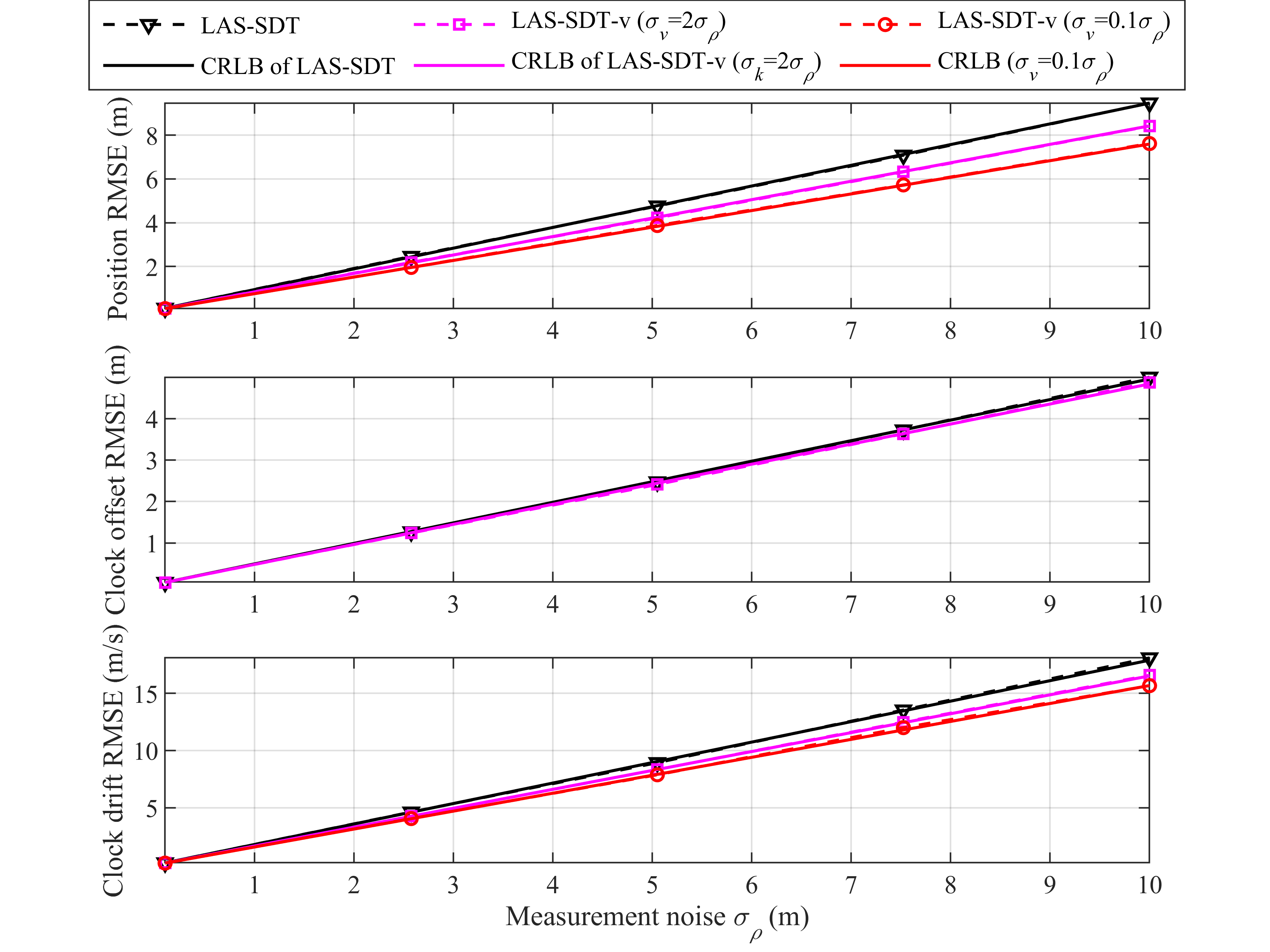} 
	\caption{Estimation error vs. measurement noise for the LAS-SDT-v. All the estimation errors of the LAS-SDT-v reach CRLB. The estimation accuracies of the LAS-SDT-v are higher than those of the LAS-SDT, and degrade with increasing aiding velocity error $\sigma_v$.}
	\label{fig:resultknownV}
\end{figure}

The estimation errors of the UD position, clock offset and clock drift are shown in Fig. \ref{fig:resultknownV}. We also depict the case with $\sigma_{v}=2\sigma_{\rho}$ for comparison. We can see that the estimation errors of the position, clock offset and clock drift from the LAS-SDT-v all reach their CRLBs. Compared with the LAS-SDT, the LAS-SDT-v has smaller errors. When the aiding velocity error increases, the estimation error approaches that of the LAS-SDT. This is consistent with the theoretical analysis in Section \ref{knownV1}. We note that the accuracy improvement of the clock offset is smaller than the improvement of the position. The reason is that the velocity has a more direct relation to the position than to the clock offset. Therefore, the velocity information contributes more on the improvement of the position accuracy.

\subsubsection{Impact of Deviated Velocity Information}
We fix the TOA measurement noise to $\sigma_{\rho}$=0.1 m, and deviate the aiding UD velocity from its true value. We vary the norm of the velocity deviation from 0 to 50 m/s with 6 steps. The direction of the velocity deviation is randomly selected from $\mathcal{U}(0,2\pi)$. The UD positions are randomly selected from the \textit{Inside Case}.

\begin{figure}
	\centering
	\includegraphics[width=0.99\linewidth]{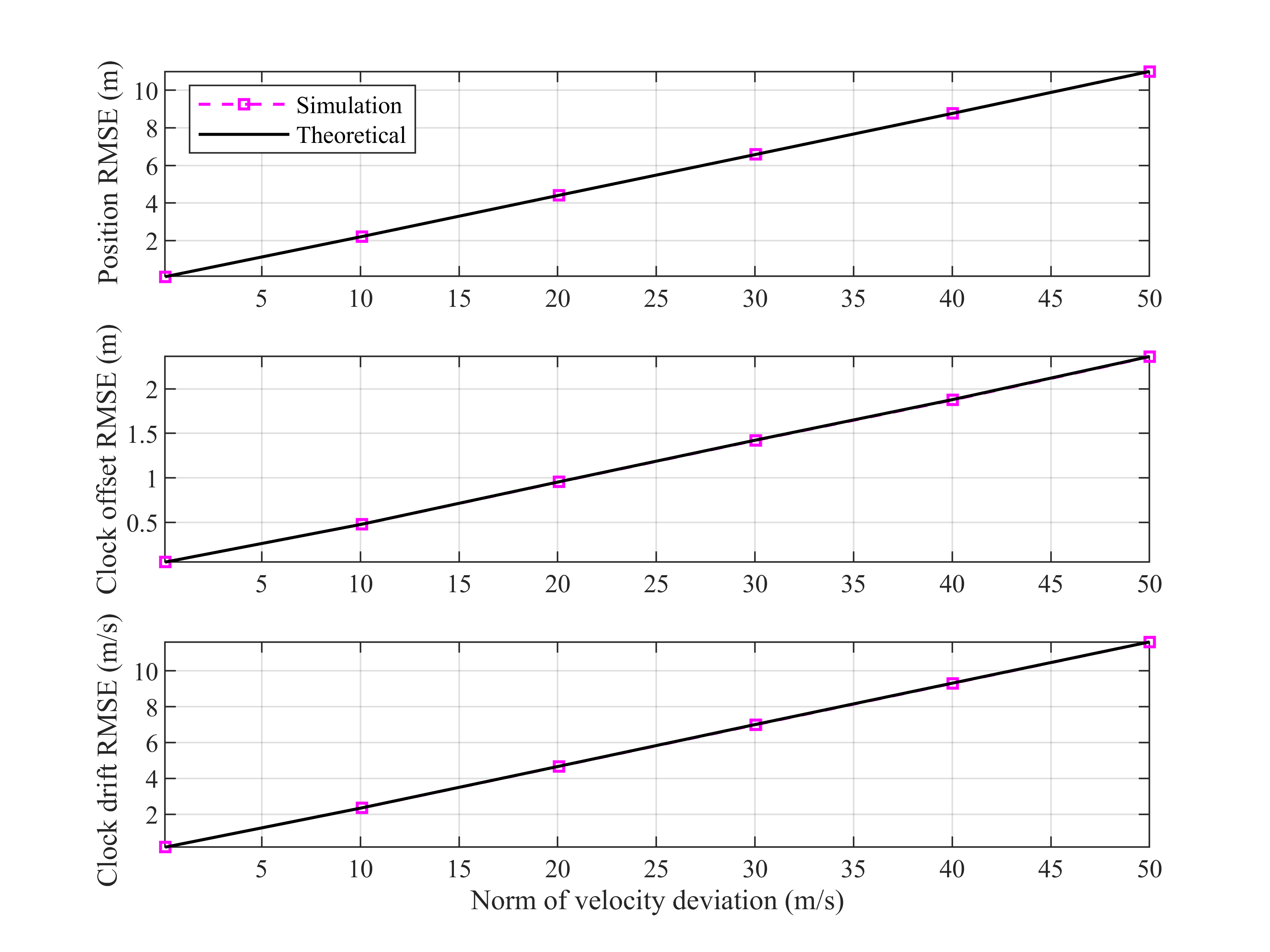} 
	\caption{Estimation error vs. norm of UD velocity deviation for LAS-SDT-v. The estimation errors increase with growing velocity deviation. The estimation errors from simulations all match the theoretical analysis.}
	\label{fig:DeviateV}
\end{figure}

We plot the estimation RMSEs in Fig. \ref{fig:DeviateV}. As can be seen, the estimation errors of position, clock offset and clock drift all increase when the deviated UD velocity becomes larger. The theoretical curves are obtained based on \eqref{eq:RMSEdV}. The figure shows that the RMSEs from the numerical simulations all match the theoretical analysis.

\subsection{LAS Performance with UD Clock Drift Aiding (LAS-SDT-k)} \label{knownO}
\subsubsection{LAS with Aiding Clock Drift Subject to Random Error}
For the LAS-SDT-k, we set the aiding UD clock drift to the true value added with a zero-mean Gaussian random error. The UD positions are randomly selected from the \textit{Inside Case}. The STD of the aiding clock drift error is set to $\sigma_k=0.5\sigma_{\rho}$ in m/s. The other simulation settings are the same as Section \ref{simLASSTOAD}.

\begin{figure}
	\centering
	\includegraphics[width=0.99\linewidth]{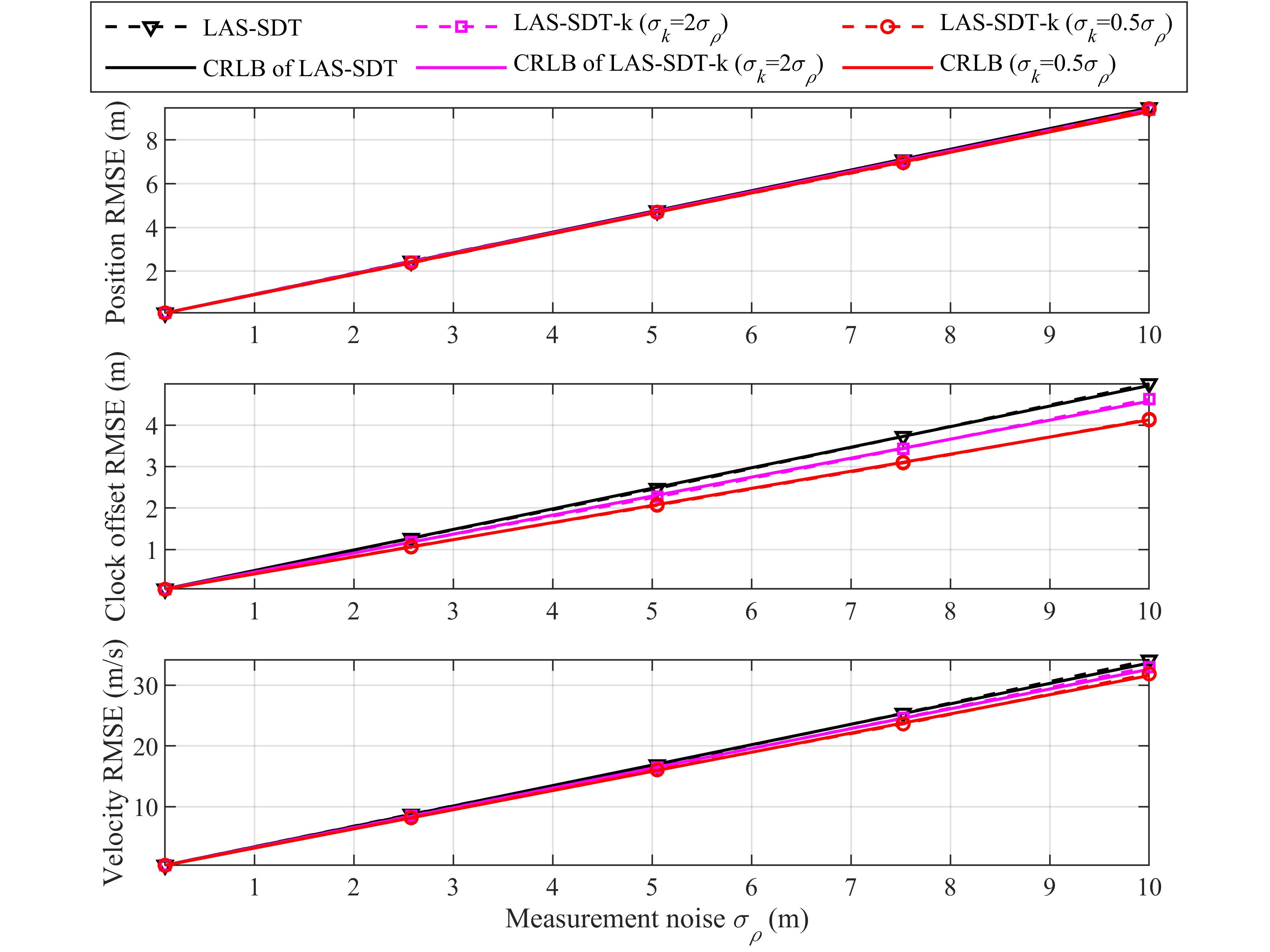} 
	\caption{Estimation error vs. measurement noise for LAS-SDT-k. All the estimation errors of the LAS-SDT-k reach CRLB and are smaller than those of the LAS-SDT, and will increase when the aiding clock drift error $\sigma_k$ grows.}
	\label{fig:RMSEvsnoiseKO}
\end{figure}

The estimation errors of the UD position, clock offset and velocity are shown in Fig. \ref{fig:RMSEvsnoiseKO}. We plot the case with $\sigma_k=2\sigma_{\rho}$ for comparison. As can be seen from the figure, the estimation errors of the position, clock offset and velocity of LAS-SDT-k all reach the CRLB and are smaller than those of the LAS-SDT. When the aiding clock drift error increases, the estimation error approaches that of the LAS-SDT.  We note that the position error of the LAS-SDT-k is only slightly smaller than that of the LAS-SDT. Since the aiding clock drift information is more related to the clock offset, the estimation error of the clock offset is more improved than the position estimation. The results shown in the figure corroborate the theoretical analysis in Section \ref{knownO1}.

\subsubsection{Impact of Deviated Clock Drift Information}
We set the TOA measurement noise to $\sigma_{\rho}=0.1$ m. The absolute value of the clock drift deviation varies from 0 to 0.2 ppm, which equals to an LOS speed range from 0 m/s to 60 m/s in a TDBS using RF signals. The UD positions are randomly selected from the \textit{Inside Case}.

\begin{figure}
	\centering
	\includegraphics[width=0.99\linewidth]{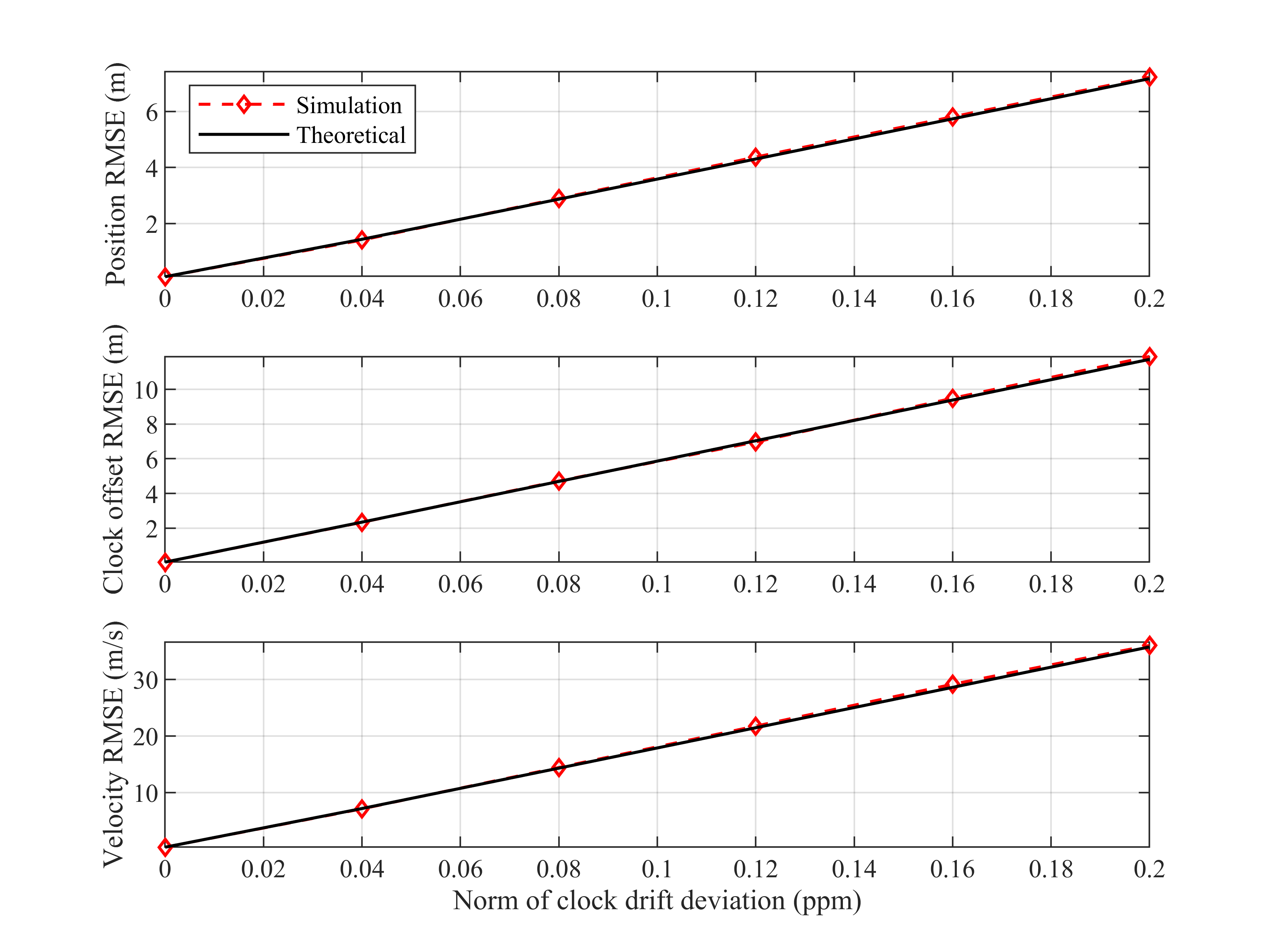} 
	\caption{Estimation error vs. UD clock drift deviation for LAS-SDT-k. The estimation errors increase with growing clock drift deviation. The estimation errors from simulations are consistent with the theoretical analysis.}
	\label{fig:DeviateOme}
\end{figure}

The estimation errors are plotted in Fig. \ref{fig:DeviateOme}. It shows that the estimation errors of position, clock offset and velocity all increase when the aiding clock drift deviates from the true value. The theoretical curves are obtained based on \eqref{eq:RMSEdome}. The results show that the estimation errors from numerical simulations are consistent with the theoretical analysis.

\subsection{Computational Complexity}
We can see from Algorithm 1 that the iterative algorithm for LAS-SDT has the same operations in each iteration. In each iteration, the major operations are the matrix multiplication and inverse given by \eqref{eq:leastsquare}. The complexity of these operations is on the order of $K^3$ \cite{quintana2001note}, where $K$ is the dimension of the matrix. The total complexity is on the order of $LK^3$, where $L$ is the number of iterations.

We also record the run time of the LAS-SDT algorithm in the numerical simulation. The computation platform we use is Matlab R2019b on a PC with Intel Core i5-10600K CPU @ 4.1 GHz and 32-GB RAM. We force the number of iterations in the algorithm to be 10, which is more than sufficient to obtain a correct solution, and record the total computation time of 5,000 runs. We have found that the average computation time for each run is only 0.90 ms. Such a low complexity is suitable for real-time operation on size, weight and power (SWaP) constrained devices such as IoT systems, miniature drones and robotics.

\section{Conclusion}
In this paper, we develop a new LAS-SDT method to exploit the sequential Doppler shifts, which can be measured in the TDBS, to obtain higher estimation accuracies for the position, velocity, clock offset and clock drift, than the commonly used TOA-only methods. The new LAS-SDT method is applicable for real-world scenarios, where moving UDs always have oscillator frequency offsets.


Particularly, we formulate the LAS problem as an ML estimator and present the iterative algorithm, which achieves optimal LAS estimation and is suitable for SWaP devices. We further develop two LAS-SDT variants for two special scenarios when additional prior information is available: (i) LAS-SDT-v for the case with UD velocity aiding, and (ii) LAS-SDT-k for the case with UD clock drift aiding. Theoretical analyses confirm our expectation that the LAS-SDT has higher LAS accuracy than the conventional TOA-only method, and the LAS-SDT-v and LAS-SDT-k achieve additional gain with the prior knowledge on UD velocity or clock drift. 

Numerical results show that the estimation error reaches the CRLB. The LAS estimation accuracy of the new LAS-SDT is significantly higher than that of the conventional LSPM-UVD method using TOA measurements only. All the numerical results are consistent with the theoretical analyses.

\appendices
\section{Proof of Remark 1}
\label{Appendix1}
Based on the definitions of $\bm{W}$ in \eqref{eq:weightmatall} and $\bm{G}$ in \eqref{eq:GUderivative}, we partition them into TOA-related and Doppler-related sub-matrices as
\begin{align}\label{eq:partitionWG}
	\bm{W}=\left[
	\begin{matrix}
		\bm{W}_{d}& \bm{O}_{M} \\
		\bm{O}_{M} & \bm{W}_{\rho}
	\end{matrix}
	\right], \;
	\bm{G}=\left[
	\begin{matrix}
		\bm{G}_{{d}} \\
		\bm{G}_{{\rho}}
	\end{matrix}
	\right],
\end{align}
where
\begin{align}\label{eq:subW}
	\bm{W}_{d}&=  \mathrm{diag}\left(\frac{1}{\sigma_{d_1}^2},\cdots,\frac{1}{\sigma_{d_M}^2}\right),\nonumber\\
	\bm{W}_{\rho}&=  \mathrm{diag}\left(\frac{1}{\sigma_{\rho_1}^2},\cdots,\frac{1}{\sigma_{\rho_M}^2}\right),
\end{align}
\begin{align}\label{eq:subG}
	[{\bm{G}}_{{d}}]_{i,:} &=
	\left[\frac{\left[\partial h\right]_{i}}{\partial \boldsymbol{p}},0,\frac{\left[\partial h\right]_{i}}{\partial \boldsymbol{v}},1 \right],i=1,\cdots,M,\nonumber\\
	[{\bm{G}}_{{\rho}}]_{i,:} &=
	\left[-{\boldsymbol{e}}_{i}^T,1,-{\boldsymbol{e}}_{i}^T\Delta t_i,\Delta t_i \right], i=1,\cdots,M.
\end{align}

The FIM for the LAS-SDT in \eqref{eq:UFIMvsG} is re-written as
\begin{align}\label{eq:UFIMvsG1}
	\mathcal{F} &= 
	\begin{bmatrix}
		\bm{G}_{d}^T &\bm{G}_{{\rho}}^T
	\end{bmatrix}
	\begin{bmatrix}
		\bm{W}_{d}& \bm{O}_{M} \\
		\bm{O}_{M} & \bm{W}_{\rho}
	\end{bmatrix}
	\begin{bmatrix}
		\bm{G}_{{d}}\\
		\bm{G}_{{\rho}} 
	\end{bmatrix}\nonumber\\
	&=\bm{G}_{{d}}^T\bm{W}_{d}\bm{G}_{{d}}+\bm{G}_{{\rho}}^T\bm{W}_{\rho}\bm{G}_{{\rho}}
	\text{.}
\end{align}

Note that the matrix $\bm{G}_{{\rho}}$ is the design matrix for the LSPM-UVD in \cite{zhao2020optloc}. We denote the FIM of the LSPM-UVD by $\mathcal{F}_{\text{LSPM-UVD}}$ and
\begin{align}\label{eq:FIMUVD}
	\mathcal{F}_{\text{LSPM-UVD}}  
	=\bm{G}_{{\rho}}^T\bm{W}_{\rho}\bm{G}_{{\rho}}
	\text{.}
\end{align}

We know that the matrix $\bm{G}_{{d}}^T\bm{W}_{d}\bm{G}_{{d}}$ is positive-definite. Then, we have
\begin{align}\label{eq:FIMcompare}
	\mathcal{F} \succ \mathcal{F}_{\text{LSPM-UVD}}  
	\text{.}
\end{align}

The CRLB is the diagonal elements in the inverse FIM. Therefore, the CRLBs of the LAS-SDT and the LSPM-UVD have the relation as
\begin{equation} \label{eq:CRLBcompare}
	\mathsf{CRLB}<\mathsf{CRLB}_{\text{LSPM-UVD}} \text{.}
\end{equation}

Thus, we have proven that the LAS accuracy of the proposed LAS-SDT is higher than that of the conventional LSPM-UVD.

\section{Proof of Remark 2}
\label{Appendix2}
We re-write the design matrix for the LAS-SDT-v, i.e., $\bm{G}_v$ as
\begin{align} 
	[{\bm{G}}_v]_{i,:} =\left[\begin{matrix}
			{\bm{G}}\\
			\begin{matrix}
				\bm{O}_{N\times(N+1)}&\bm{I}_N&\bm{0}_{N}
			\end{matrix}
		\end{matrix}\right].
\end{align}

We define
\begin{align}
	\bm{\Lambda}\triangleq\begin{bmatrix}
		\bm{O}_{N\times(N+1)}&\bm{I}_N&\bm{0}_{N}
	\end{bmatrix}.
\end{align}

The FIM for the LAS-SDT-v in \eqref{eq:KVCRLBFisher} becomes
\begin{align}\label{eq:FIMvnew}
	\mathcal{F}_v &= \bm{G}_v^T\bm{W}_v\bm{G}_v=\begin{bmatrix}
		\bm{G}^T&\bm{\Lambda}^T
	\end{bmatrix}
	\begin{bmatrix}
		\bm{W}&\\
		&\bm{\Sigma}_v^{-1}
	\end{bmatrix}
\begin{bmatrix}
	\bm{G}\\
	\bm{\Lambda}
\end{bmatrix}\nonumber\\
&=\bm{G}^T\bm{W}\bm{G}+\bm{\Lambda}^T\bm{\Sigma}_v^{-1}\bm{\Lambda}.
\end{align}

Note that $\bm{\Lambda}^T\bm{\Sigma}_v^{-1}\bm{\Lambda}$ is positive semi-definite, and $\bm{G}^T\bm{W}\bm{G}$ is positive definite when there are sufficient number of ANs and a proper geometry. Therefore,
\begin{align}\label{eq:FIMvFIM}
	\bm{G}^T\bm{W}\bm{G}+\bm{\Lambda}^T\bm{\Sigma}_v^{-1}\bm{\Lambda}\succeq \bm{G}^T\bm{W}\bm{G}.
\end{align}

We apply inverse on both sides of \eqref{eq:FIMvFIM} and come to
\begin{align}\label{eq:velgain}
	\left(\bm{G}^T\bm{W}\bm{G}+\bm{\Lambda}^T\bm{\Sigma}_v^{-1}\bm{\Lambda}\right)^{-1}\preceq \left(\bm{G}^T\bm{W}\bm{G}\right)^{-1}.
\end{align}

We can see from \eqref{eq:velgain} that the achievable estimation error of the LAS-SDT-v is smaller than that of the LAS-SDT, and the performance gain of the LAS-SDT-v is the term $\bm{\Lambda}^T\bm{\Sigma}_v^{-1}\bm{\Lambda}$. Moreover, when the velocity error grows, i.e., $\bm{\Sigma}_v\rightarrow \infty$, the estimation error of the LAS-SDT-v method will increase and approach that of the LAS-SDT. Remark 2 is proven.

\section{Derivation of Remark 3}
\label{Appendix3}

According to (\ref{eq:dPvsdV}), we have the estimation bias as
\begin{equation}\label{eq:squaredbias}
	\Vert\bar{\boldsymbol{\mu}}_{v}\Vert^2=\bar{\boldsymbol{r}}_v^T\bm{S}_1^T\bm{S}_1\bar{\boldsymbol{r}}_v \text{,}
\end{equation}
where $\bm{S}_1=({\bm{G}}_v^T\bm{W}_v{\bm{G}}_v)^{-1}{\bm{G}}_v^T\bm{W}_v$.


We rewrite $\bar{\boldsymbol{r}}_{v}$ given by \eqref{eq:resdV} as
\begin{align} \label{eq:rexpansion}
	\bar{\boldsymbol{r}}_{v}=\bm{S}_2\Delta\boldsymbol{v} \text{.}
\end{align}
where the deviated aiding velocity $\Delta \boldsymbol{v}=\boldsymbol{v}-\bar{\boldsymbol{v}}$.

By plugging (\ref{eq:rexpansion}) into (\ref{eq:squaredbias}), we come to
\begin{align}\label{eq:squarebias1}
	\Vert\bar{\boldsymbol{\mu}}_{v}\Vert^2=\Delta\boldsymbol{v}^T\bm{S}\Delta\boldsymbol{v}\text{,}
\end{align}
where $\bm{S}=\bm{S}_2^T\bm{S}_1^T\bm{S}_1\bm{S}_2$.

We have obtained the relation between the estimation bias and the deviated aiding velocity as given by \eqref{eq:squarebias1}. Further more, note that $\bm{S}$ is positive definite. Therefore, there must be a positive scalar $\beta$ to make the matrix $\bm{S}-\beta\bm{I}$ positive semi-definite. Therefore, we come to
\begin{align}\label{eq:squareerror2}
	\Vert\bar{\boldsymbol{\mu}}_{v}\Vert^2= \beta\Delta\boldsymbol{v}^T\Delta\boldsymbol{v} + \Delta\boldsymbol{v}^T\left(\bm{S}-\beta\bm{I}\right)\Delta\boldsymbol{v} \geq\beta\Vert \Delta\boldsymbol{v}\Vert^2
\end{align}

We can see from (\ref{eq:squareerror2}) that the estimation bias is growing when the speed aiding deviation increases.


%


\ifCLASSOPTIONcaptionsoff
\newpage
\fi



\bibliographystyle{IEEEtran}
\bibliography{IEEEabrv,paper}

\end{document}